\documentclass[preprint,nofootinbib,aps,superscriptaddress,eqsecnum]{revtex4-1}
\pdfoutput=1
\usepackage[colorlinks=true,citecolor=blue,linkcolor=blue,breaklinks=true]{hyperref}
\usepackage{graphicx}
\usepackage{mathrsfs}
\usepackage{soul}
\usepackage{pifont}
\usepackage[utf8]{inputenc}
\usepackage{enumerate}
\usepackage{amsmath,amssymb}
\usepackage{adjustbox}
\usepackage{slashed}
\usepackage{amssymb,amsmath,subfigure,xcolor}

\newcommand{\lapprox}{%
\mathrel{%
\setbox0=\hbox{$<$}
\raise0.6ex\copy0\kern-\wd0
\lower0.65ex\hbox{$\sim$}
}}
\newcommand{\gapprox}{%
\mathrel{%
\setbox0=\hbox{$>$}
\raise0.6ex\copy0\kern-\wd0
\lower0.65ex\hbox{$\sim$}
}}

\newcommand{\ba}{\begin{array}}
\newcommand{\ea}{\end{array}}
\newcommand{\bd}{\begin{displaymath}}
\newcommand{\ed}{\end{displaymath}}
\newcommand{\beq}{\begin{equation}}
\newcommand{\eeq}{\end{equation}}
\newcommand{\bea}{\begin{eqnarray}}
\newcommand{\eea}{\end{eqnarray}}
\newcommand{\bpm}{\begin{pmatrix}}
\newcommand{\epm}{\end{pmatrix}}
\newcommand{\gbl}{g_{BL}}

\newcommand{\nn}{\nonumber}

\def\ie{ {\em i.e.,\ }}

\def\a{\alpha}

\def\k{\kappa}
\def\g{\gamma}

\def\l{\lambda}
\def\m{\mu}
\def\n{\nu}

\def\q2 {q^2}
\def\mz1 {m_{Z_1}}

\def\bt{\begin{table}}
\def\et{\end{table}}

\catcode`@=11 
\def \gsim{\mathrel{\mathpalette\@versim>}}
\def \lsim{\mathrel{\mathpalette\@versim<}}
\def \@versim#1#2{\lower0.4ex\vbox{\baselineskip\z@skip\lineskip\z@skip
     \lineskiplimit\z@\ialign{$\m@th#1\hfil##\hfil$%
     \crcr#2\crcr\sim\crcr}}}
\catcode`@=12 

\begin{document}

\title{Constraints on the Doublet Left-Right Symmetric Model\\ from Higgs data }

\author{Siddhartha Karmakar\footnote{siddhartha.karmakar$\_315$@tifr.res.in}}
\affiliation{{\small Department of Physics, Indian Institute of Technology Bombay,\\
 Powai, Mumbai-400076, India}}
\affiliation{{\small Department of Theoretical Physics, Tata Institute of Fundamental Research,\\ Colaba, Mumbai-400005, India}}
\author{Jai More\footnote{more.physics@gmail.com}}
\affiliation{{\small Department of Physics, Indian Institute of Technology Bombay,\\
 Powai, Mumbai-400076, India}}
\author{Akhila Kumar Pradhan\footnote{akhilpradhan@iitb.ac.in}}
\affiliation{{\small Department of Physics, Indian Institute of Technology Bombay,\\
 Powai, Mumbai-400076, India}}
\author{S. Uma Sankar\footnote{uma@phy.iitb.ac.in}}
\affiliation{{\small Department of Physics, Indian Institute of Technology Bombay,\\
 Powai, Mumbai-400076, India}}

\vspace{30pt}

\begin{abstract}
We study the constraints on the doublet left-right symmetric model~(DLRSM) arising due to the Higgs
data. The $SU(2)_L$ symmetry of this model is broken by three vacuum expectation values,
$\k_1$, $\k_2$ and $v_L$. Most studies of this model assume that the ratios $r = \k_2/\k_1$ and
$w = v_L/\k_1$ are very small. In this work, we study the constraints imposed on $r$ and $w$
by the Higgs data from LHC. We consider the most general scalar potential and calculate 
the masses of the CP-even scalars and the couplings of the lightest of these scalars to itself, to $W$ and $Z$ gauge bosons and to the third generation quarks. We find that there is no lower bound on either $r$
or $w$. Equating the mass of the lightest CP-even scalar to $125$\,GeV leads to an upper limit $w < 6.7$. The requirement that the Yukawa coupling of the quarks to the Higgs bidoublet of the model should be perturbative yields the upper bounds $r < 0.8$ and $w < 3.5$. The Yukawa coupling of the bottom quark to the lightest CP-even scalar strongly disfavours value of $r, w < 0.1$ and shows a marked preference for values of $w \sim \mathcal{O} (1)$.
\end{abstract}

\pacs {}
\maketitle

\section{Introduction}

Left-right symmetric models, based on the symmetry group 
$SU(2)_L \times SU(2)_R \times U(1)_{\rm B-L}$, have been attractive candidates for
physics beyond the standard model (SM)~\cite{Pati:1974yy,Mohapatra:1974gc,Mohapatra:1974hk,Senjanovic:1975rk,Senjanovic:1978ev}. In these models, the left-chiral fermions are doublets under
$SU(2)_L$ and singlets under $SU(2)_R$. The situation is reversed for the 
right-chiral fermions. The $U(1)_{\rm B-L}$ hyper-charges of the fermions are fixed 
by the modified Gell-Mann-Nishijima formula connecting them to their respective electric charges. These models have been deployed to address numerous open 
problems of particle physics, such as neutrino mass generation, dark matter, unification of forces, etc.

The scalar sector of this model requires a minimum of three different
Higgs multiplets. Fermion mass generation requires a scalar bidoublet $\Phi$,
which transforms as a doublet under both $SU(2)_L$ and $SU(2)_R$. In addition,
we need two more scalar multiplets to break the gauge symmetry of the model to
$U(1)_{\rm em}$ and to maintain the left-right symmetry of the Lagrangian.
These additional scalars can be either $SU(2)$ doublets or triplets. 
In doublet left-right symmetric models~(DLRSM), they are denoted as 
$\chi_L$ ($\chi_R$) which is a doublet~(singlet) under $SU(2)_L$ and a 
singlet~(doublet) under $SU(2)_R$. When the neutral component of $\chi_R$ 
acquires the vacuum expectation value~(\textit{vev}) $v_R$, the symmetry of the model
is broken to that of the SM. When the neutral components of $\chi_L$ and
$\Phi$ acquire \textit{vev}s, the symmetry is finally broken to $U(1)_{\rm em}$.
Since there is no evidence yet for any right-handed currents in weak interactions,
the value of $v_R$ has to be much greater than the \textit{vev}s of $\chi_L$ and $\Phi$.
If the additional scalars are $SU(2)$ triplets, the model is called 
triplet left-right symmetric model~(TLRSM). In this model, it is possible to generate
light Majorana masses for neutrinos, through both type-I and type-II see-saw
mechanisms. The constraints on the scalar sector of this model are qualitatively
different from those of DLRSM. In this work, we confine our attention to DLRSM.

The minimal form of DLRSM was originally discussed in ref.~\cite{Senjanovic:1978ev}. 
Ref.~\cite{Babu:1988qv} extended this model by adding a singly-charged $SU(2)$ singlet complex scalar so that small neutrino masses can be generated at one-loop level by the Zee mechanism~\cite{Zee:1980ai}. Such a scalar mixes with the charged scalars in DLRSM.  
The details of this mixing are constrained from the neutrino oscillation data~\cite{Babu:2020bgz}, flavour violating decays of charged leptons~\cite{FileviezPerez:2017zwm}, and collider searches for RH neutrinos~\cite{FileviezPerez:2016erl}. 

It is an interesting question to ask what the dominant source of electroweak symmetry breaking
(EWSB) in DLRSM is: whether it is the \textit{vev} $v_L$ of the doublet $\chi_L$ or the 
\textit{vev}s $\k_1$ and $\k_2$ of the bidoublet $\Phi$. Most models, based on left-right symmetry, assume the latter to be the case. By considering the bounds from perturbative unitarity of scalar and gauge boson scattering and the electroweak precision data, ref.~\cite{Bernard:2020cyi} showed that both the above possibilities are allowed. It means there is no hierarchy between $v_L$ and $\k_1$, $\k_2$. Most left-right symmetric models also assume $\k_2 \ll \k_1$. It will be interesting to see what bounds the data imposes on this hierarchy. The purpose of the current study is to analyse the pattern of EWSB in the simplest DLRSM in light of Higgs data and theoretical arguments such as perturbativity, unitarity, and boundedness from below. 

In sec.~\ref{sec:notations} we introduce our notations and discuss the key features of the different sectors of DLRSM. In sec.~\ref{sec:theoreticalbounds} we mention the bounds from perturbativity, unitarity, and boundedness from below. A convenient and minimal basis of quartic parameters have been introduced in sec.~\ref{sec:convenient}. Salient features of the model in light of Higgs data are presented in sec.~\ref{sec:higgsbounds}. This section also discusses the validity of the assumption of hierarchy in the \textit{vev}s $v_L, \k_2 \ll \k_1$ as a function of the values of the quartic couplings in the model. We make our concluding remarks in 
sec.~\ref{sec:summary}.
     
\section{The Doublet Left-Right Symmetric Model: Essentials and Notations}
\label{sec:notations}

DLRSM  is based on the gauge symmetry $SU(3)_C \times SU(2)_L \times SU(2)_R \times U(1)_{B-L}$. The fermion content of the model is given below, where the parenthesis contain the quantum numbers
of the fermions under each of the sub-groups of the gauge group,
 \begin{align}
Q_L &= \begin{pmatrix}
 u_L \\
 d_L \\
\end{pmatrix} \sim (3,2,1,1/3), \hspace{10mm}
Q_R = \begin{pmatrix}
 u_R \\
 d_R \\
\end{pmatrix} \sim (3,1,2,1/3), \nonumber\\[5pt]
L_L &= \begin{pmatrix}
 \nu_L \\
 e_L \\
\end{pmatrix} \sim (1,2,1,-1), \hspace{12mm}
L_R = \begin{pmatrix}
 \nu_R \\
 e_R \\
\end{pmatrix}\sim (1,1,2,-1).
\label{eq:fermion}
\end{align}
For any left-right symmetric model, parity is truly manifest as a symmetry, making the Lagrangian symmetric under transformations $Q_L \leftrightarrow Q_R$, $L_L \leftrightarrow L_R$. Note that the model necessarily contains a right-handed neutrino $\nu_R$, which is required to complete the lepton doublet.
Both $\n_L$ and $\n_R$ receive Majorana masses if DLRSM  is extended with additional scalars~\cite{FileviezPerez:2017zwm} or fermions~\cite{Brdar:2018sbk}. In our work, we do not include these additional fields.

\subsection{The scalar potential}

The scalar sector consists of a complex bidoublet and two doublets with the following charges under the aforementioned gauge group  
\bea
 \Phi = \begin{pmatrix}
 \phi_1^0  & \phi_2^+  \\
 \phi_1^-  & \phi_2^0 \\
\end{pmatrix} \sim (1,2,2,0), \hspace{5pt}
\chi_L  = \begin{pmatrix}
\chi_L^+ \\
 \chi_L^0 \\
\end{pmatrix} \sim (1,2,1,1), \hspace{5pt}
\chi_R = \begin{pmatrix}
 \chi_R^+ \\
 \chi_R^0 \\
\end{pmatrix} \sim (1,1,2,1)\,\,, \nn
\eea
with the {\it vev} structure 
\begin{align}
  \langle \Phi \rangle =  \frac{1}{\sqrt{2}}\begin{pmatrix}
 \kappa_1 & 0 \\
 0 &\hspace{3mm} \ \kappa_2 \\
\end{pmatrix} \, , \hspace{7mm}
\langle \chi_L \rangle =
\frac{1}{\sqrt{2}}\begin{pmatrix}
 0 \\
  v_L \\
\end{pmatrix} \, , \hspace{7mm}
\langle \chi_R \rangle =
\frac{1}{\sqrt{2}} \begin{pmatrix}
0 \\
 v_R \\
\end{pmatrix} \, .
\label{eq:vev}
\end{align} 
The spontaneous breaking of $SU(2)_R \times U(1)_{B-L} \rightarrow U(1)_Y$ is driven by the {\it vev} of the doublet $\chi_R$ whereas the EWSB is triggered by the three \textit{vev}s $\k_1$, $\k_2$, and $v_L$.
The last three \textit{vev}s are constrained by $\k_1^2+\k_2^2+v_L^2 = v^2$, where $v = 246$\,GeV. For later use, it is convenient to introduce the ratios $r = \k_2/\k_1$ and $w= v_L/\k_1$. 
The \textit{vev}s of the scalars must have the hierarchy $v_R \gg v$, which ensures that the gauge bosons of $SU(2)_R$ are much heavier than the weak gauge bosons. The Yukawa couplings of the bidoublet with fermions lead to fermion masses and mixings.  The details of these couplings will be discussed later in sec.~\ref{sec:fermionsector}.

The most general, CP-conserving, renormalizable Higgs potential involving 
$\Phi, \chi_L$ and $\chi_R$ fields is given by
\bea 
V  &=& V_2 + V_3 + V_4,\nn\\
V_2  &=& -\mu _1^2\text{Tr}(\Phi^{\dagger}\Phi) - \mu_2^2\ [\text{Tr}(\tilde{\Phi}\Phi^{\dagger})+ \text{Tr}(\tilde{\Phi}^{\dagger} \Phi)] - \mu_3^2\ [\chi_L^{\dagger} \chi_L + \chi_R^{\dagger} \chi_R] \,\, , \nn\\
V_3 &=& \mu_4\  [\chi_L^{\dagger} \Phi \chi_R + \chi_R^{\dagger} \Phi^{\dagger} \chi_L] + \mu_5\  [\chi_L^{\dagger} \tilde{\Phi} \chi_R + \chi_R^{\dagger}\tilde{\Phi}^{\dagger}\chi_L ]\,\, , \nn\\
V_4 &=& \lambda_1\text{Tr}(\Phi^{\dagger}\Phi)^2 + \lambda_2\ [ \text{Tr}(\tilde{\Phi} \Phi^{\dagger})^2
 + \text{Tr}(\tilde{\Phi}^{\dagger} \Phi)^2 ]  + \lambda_3\text{Tr}(\tilde{\Phi} \Phi^{\dagger}) \, \text{Tr}(\tilde{\Phi}^{\dagger} \Phi)  \nn\\
 &&+ \lambda_4\text{Tr}(\Phi^{\dagger}\Phi) \, [\text{Tr}(\tilde{\Phi}\Phi^{\dagger})+ \text{Tr}(\tilde{\Phi}^{\dagger}\Phi)]  + \rho_1\  [(\chi_L^{\dagger} \chi_L )^2 + (\chi_R^{\dagger} \chi_R )^2]
 + \rho_2\  \chi_L^{\dagger} \chi_L \chi_R^{\dagger}\chi_R \nn\\
 &&+ \alpha_1\text{Tr}(\Phi^{\dagger} \Phi ) [\chi_L^{\dagger}\chi_L + \chi_R^{\dagger}\chi_R ]
 + \Big\{ \alpha_2 \ [\chi_L^{\dagger} \chi_L  \text{Tr}(\tilde{\Phi} \Phi^{\dagger} ) + \chi_R^{\dagger} \chi_R  \text{Tr}(\tilde{\Phi}^{\dagger} \Phi )] + {\rm h.c.} \Big\} \nn\\
 &&+ \alpha_3\ [\chi_L^{\dagger}\
 \Phi \Phi^{\dagger}\chi_L + \chi_R^{\dagger} \Phi^{\dagger} \Phi  \chi_R  ] 
 + \alpha_4\ [\chi_L^{\dagger}\
 \tilde{\Phi} \tilde{\Phi}^{\dagger}\chi_L + \chi_R^{\dagger} \tilde{\Phi}^{\dagger} \tilde{\Phi}  \chi_R  ]\,\,.
 \label{eq:scalarpotential}
\eea
Here all the couplings can be made real by appropriate field redefinitions.

The following conditions minimize the potential
\begin{equation}
     \frac{\partial V}{\partial \kappa_1} = \frac{\partial V}{\partial \kappa_2} = \frac{\partial V}{\partial v_L} = \frac{\partial V}{\partial v_R} =  0\,\,.
\end{equation}
We utilize these conditions to replace the four
mass parameters $\mu_1^2, \mu_2^2, \mu_3^2$ and $\mu_5$ by 
the \textit{vev}s and quartic couplings, 
\bea 
\mu_1^2 &=& \frac{1}{2(r^2 -1)}\Bigg(\k_1^2 \Big(w^2((r^2-1)\a_1 + r^2 \a_3 - \a_4) + 2 (r^2 -1)((r^2 + 1)\l_1 + 2 r \l_4) \Big) \nn\\
&& + 2 \sqrt{2} r v_R w \m_4 + v_R^2 \Big((r^2 -1)\a_1 + r^2 \a_3 - \a_4  + 2 w^2 \rho_{12}\Big) \Bigg) \,\,, \nn\\
\mu_2^2 &=&  \frac{1}{4(r^2 -1)}\Bigg( \k_1^2 \Big(w^2 (r^2-1) \a_2 - w^2 r \a_{34} + 2 (r^2 -1)(2 r \l_{23} + (r^2+1)\l_4) \Big)\nn\\
&& - \sqrt{2}(r^2+1) v_R w \m_4 + v_R^2 \Big((r^2-1)\a_2 - r \a_{34} - 2 w^2 \rho_{12} \Big) \Bigg)\,\,, \nn\\
\m_3^2 & = & \frac{1}{2} \k_1^2 ((r^2+1)\a_1 + 2 r \a_2 + r^2 \a_3 + \a_4 + 2 w^2 \rho_{1}) + v_R^2 \rho_1\,\,, \nn\\
\m_5 &=& - r \m_4 - \sqrt{2} v_R w \rho_{12}   \,\,  .
\label{eq:minimisationcondition}
\eea
Here $\rho_{12}=\rho_2/2 -\rho_1$, $\alpha_{34}=\alpha_3-\alpha_4$, and $\l_{23} = 2 \l_2 + \l_3$. From the definition of $r$ and $w$, we note that $\k_1^2 (1 + r^2 + w^2) = v^2$. That is, the value of $\k_1$ is fixed for a given $r$ and $w$. Thus the parameters of the DLRSM scalar sector are
\bea 
\{\l_{1,2,3,4}, \a_{1,2,3,4}, \rho_{1,2}, \m_4, r, w, v_R \}\,\, .
\eea
Among these, the only  dimensionful parameter is $\m_4$. For most of the following discussions, we will either set it to zero or restrict it to $\mu_4 \lesssim v_R$. 

\subsection{The gauge sector}
\label{sec:gaugesector}

The charged gauge boson mass matrix in this model is
\bea 
\mathcal{L}_{mass} \supset 
\begin{pmatrix}
W_L^+ & W_R^+
\end{pmatrix}
\begin{pmatrix}
\frac{g_L^2}{4}(v_L^2+\k_1^2+\k_2^2) & -\frac{1}{2} g_L g_R \k_1 \k_2 \\
-\frac{1}{2} g_L g_R \k_1 \k_2  & \frac{g_R^2}{4}(v_R^2+\k_1^2+\k_2^2)
\end{pmatrix}
\begin{pmatrix}
 W_L^- \\
 W_R^- \\
\end{pmatrix}   .
\eea
Here $g_{L(R)}$ are the gauge couplings associated with $SU(2)_{L(R)}$.   The physical charged gauge bosons have masses
\bea 
m^2_{W_{1,2}} = \frac{1}{8} \Big(g_L^2 v^2 + g_R^2 V^2   \mp \sqrt{(g_L^2 v^2 - g_R^2 V^2)^2 + 16 g_L^2 g_R^2 \k_1^2 \k_2^2 } \Big) \,\,\,,
\label{Wmasses}
\eea
where $W_1^{\pm}$ is close to the SM $W^{\pm}$ boson and $W_2$ has mass of the order of $v_R$. The mixing between $W_{L}^{\pm}$ and $W_R^{\pm}$ given by 
\bea 
\begin{pmatrix}
 W_L^\pm \\
 W_R^\pm \\
\end{pmatrix} = 
\begin{pmatrix}
 \cos \xi & -\sin \xi \\
  \sin \xi & \cos \xi
\end{pmatrix}      
\begin{pmatrix}
 W_1^\pm \\
 W_2^\pm \\
\end{pmatrix}   ,  \nn
\eea
where 
\bea 
\tan \xi =  \frac{4 g_L g_R \k_1 \k_2}{g_L^2 v^2- g_R^2 V^2 - \sqrt{(g_L^2 v^2- g_R^2 V^2)^2 + 16 g_L^2 g_R^2 \k_1^2 \k_2^2}} \,\,\,.
\label{eq:weigenvec}
\eea
Here, we adhere to the definitions, $v^2 = \k_1^2 + \k_2^2 + v_L^2$ and $V^2 = \k_1^2 + \k_2^2 + v_R^2$. 
In the limit $\kappa_1, \kappa_2,  v_L \ll v_R$ and $g_L = g_R$, the mixing angle becomes $\xi \simeq - 2\k_1 \k_2/v_R^2$.

In a similar fashion, the neutral gauge boson mass matrix can be written as
\bea 
\mathcal{L}_{mass} \supset \frac{1}{8}\begin{pmatrix}
W^3_{L\m} & W^3_{R\m} & B_{\m} 
\end{pmatrix} 
\begin{pmatrix}
g_L^2 v^2 & -g_L g_R \k_+^2 & -g_L g_{BL} v_L^2 \\
 & g_R^2 V^2 & -g_R g_{BL} v_R^2 \\
 & & g_{BL}^2 (v_L^2 + v_R^2) 
\end{pmatrix}
\begin{pmatrix}
W^3_{L\m} \\ W^3_{R\m} \\ B_{\m} 
\end{pmatrix} \,\,, \nn\\
\eea
with the mass eigenvalues 
\bea
m^2_{Z_1,Z_2} &=& \frac{1}{8}\Big(g_L^2 v^2 + g_R^2 V^2 + g_{BL}^2 (v_L^2 + v_R^2) \nn\\
&&\mp \sqrt{(g_L^2 v^2 + g_R^2 V^2 + g_{BL}^2 (v_L^2 + v_R^2))^2 + 4 (g_L^2 g_R^2 + g_L^2 g_{BL}^2 + g_R^2 g_{BL}^2)(\k_+^4 - v^2 V^2) } \Big) \,\,,\nn\\
\label{Zmasses}
\eea
where $\k_+^2 = \k_1^2 + \k_2^2$ and $g_{BL}$ is the gauge coupling of $U(1)_{B-L}$. The lighter boson $Z_1$ is close to the SM $Z$ boson and
the heavier boson $Z_2$ has a mass of order $v_R$. The exact expressions for the mass 
eigenstates in terms of the gauge eigenstates are given in Appendix \ref{appendix:heavyhiggs}.

In the limit $\k_1, \k_2, v_L \ll v_R$ and $g_L = g_R = g$ the mixing reduces to~\cite{Dev:2016dja} 
\bea 
\bpm A_{\m} \\ Z_{1\mu}  \\ Z_{2\m} \epm =
\bpm 
s_W & c_W s_Y & c_W c_Y \\  -c_W & s_W s_Y & s_W c_Y \\ 0 & c_Y & s_Y
\epm
  \bpm W^3_{L\m} \\ W^3_{R\m} \\ B_{\m}  \epm ,
\eea 
where
\bea 
s_W &\equiv & \sin \theta_W = \frac{\gbl}{\sqrt{g^2 + 2 \gbl^2 }}\,\,,\,\,\,\,\,
c_W \equiv  \cos \theta_W = \sqrt{\frac{g^2 + \gbl^2}{g^2 + 2 \gbl^2}}\,\,,\nn\\
s_Y &\equiv &  \sin \theta_Y = \frac{\gbl}{\sqrt{g^2 + \gbl^2}}\,\,,\,\,\,\,\,
c_Y \equiv  \cos \theta_Y = \frac{g}{\sqrt{g^2 + \gbl^2}}\,\,. 
\eea
In the remainder of the paper we will assume manifest LR symmetry in the gauge sector leading to $g_L = g_R = g$.

If we retain only $v_R^2$ terms in eqs.~\eqref{Wmasses} and~\eqref{Zmasses},
we find $m_{W_2} = g v_R/2$  and $m_{Z_2} = m_{W_2}/\cos \theta_Y$. The present
experimental lower limits on charged heavy gauge boson is $m_{W^\prime} \geq 6$\, TeV~\cite{ATLAS:2019lsy}, obtained under the assumption that fermions couple 
to $W^\prime$ and to the usual $W$ with the same strength. This assumption is certainly 
valid for manifest DLRSM. Since $v_R \approx 2 m_{W_2}/g$, we must have $v_R \sim 20$\, TeV 
to satisfy the experimental lower bound on $m_{W_2}$. The experimental lower bound
on neutral heavy gauge boson is $m_{Z^\prime} \geq 5.1$\, TeV~\cite{ATLAS:2019erb,CMS:2021ctt}, which again is obtained under the  
assumption that the fermions couple to $Z^\prime$ and to the usual $Z$ with the same strength.  Such an assumption 
is approximately true for manifest DLRSM.
Since the predicted value of $m_{Z_2}$ is greater than the predicted value of $m_{W_2}$, 
the lower bound on $m_{Z_2}$ is automatically satisfied if the lower bound on $m_{W_2}$ is satisfied. Our numerical calculations give the values $m_{W_2} = 6.5$\, TeV and $m_{Z_2} = 7.7$\, TeV for $v_R =
20$\, TeV. When the $SU(2)_L$ breaking \textit{vev}s, $\k_1, \k_2$
and $v_L$ are varied subject to the constraint $\k_1^2+\k_2^2+v_L^2
= v^2$, the changes in $m_{W_2}$ and $m_{Z_2}$ are less than $1\%$.

\subsection{CP-even neutral scalars}
\label{subsection:neutral scalar}

The experiments at LHC have measured the mass of the 
Higgs boson and its couplings to gauge bosons and fermions. They have also
set lower limits on masses of heavy neutral and charged scalars.
The aim of this work is to utilize the Higgs boson measurements to constrain
the parameters of DLRSM. So we study the CP-even neutral scalars of DLRSM in detail.
For completeness, we have included the masses of CP-odd neutral 
scalars and charged scalars in Appendix~\ref{appendix:heavyhiggs}.

The gauge and physical bases for the CP-even neutral scalars are  
$X =( \phi_{1 r}^{0} , \phi_{2 r}^{0}, \chi_{L r}^{0}, \chi_{R r}^{0} )$ and 
$X_\text{ph} = (h, H_1, H_2, H_3)$ respectively. Here $h$ is the lightest 
CP-even scalar with mass of the order of $v$ and $H_i \, (i=1,2,3)$ are heavier scalars with masses of the order of $v_R$. It is expected that the properties of $h$ will be very similar to those of SM Higgs boson.

The mass matrix for scalars in the gauge basis is denoted by $M^2$.
It can be decomposed in powers of $v_R$ which enables the use of perturbation theory due to the hierarchy of scales $\k_1, \k_2, v_L \ll v_R$, 
\bea
    M^2= \Big(M^{2\,(0)} v_R^2 + M^{2\,(1)} v_R + M^{2\,(2)}\Big) \,\,\, ,
  \label{NSMMinGB}
\eea
where $M^{2\,(0)}$, $M^{2\,(1)}$ and $M^{2\,(2)}$ are three symmetric matrices.
The eigenvalues of this matrix must be positive, because they represent the masses
of the scalars.
The explicit forms of these symmetric components are 
\bea 
\big(M^{2\,(2)}\big)_{11} &=& \frac{1}{2}\k_1^2 (w^2(\a_1+\a_4) + 2 (3+r^2)\l_1 + 4 r (r \l_{23}+ 3 \l_4))\,\,, \nn\\
\big(M^{2\,(2)}\big)_{12} &=& \frac{1}{2}\k_1^2 (w^2 \a_2 + 4 r  (\l_1 + 2 \l_{23}) + 6 (r^2+1) \l_4)\,\,, \nn\\
\big(M^{2\,(2)}\big)_{13} &=& w \k_1^2 (\a_1 + r \a_2 + \a_4) \,\,, \,\,\,\,\,\,\big(M^{2\,(0)}\big)_{14} = -\frac{ r w \k_1 \m_4}{\sqrt{2}}\,\,, \nn\\
\big(M^{2\,(2)}\big)_{22} &=& \frac{1}{2}\k_1^2 (w^2(\a_1+\a_3) + 2 (3r^2+1)\l_1 + 4  ( \l_{23}+ 3 r \l_4))\,\,, \nn\\
\big(M^{2\,(2)}\big)_{23} &=& w \k_1^2 (\a_2 + r (\a_1+ \a_3))\,\,, \,\,\big(M^{2\,(0)}\big)_{24} = \frac{w \k_1 \mu_4}{\sqrt{2}}\,\,,\nn \\
\big(M^{2\,(2)}\big)_{33} &=& \frac{1}{2} \k_1^2 ((r^2+1)\a_1 + 2 r \a_2 + r^2 \a_3 + \a_4 + 6 w^2 \rho_1)\,\,,\,\,\,\,\,\big(M^{2\,(2)}\big)_{34} = 0 \,\,, \nn\\
\big(M^{2\,(2)}\big)_{44} &=& \frac{1}{2} \k_1^2 ((r^2+1)\a_1 + 2 r \a_2 + r^2 \a_3 + \a_4 + w^2 \rho_2)\,\,\,,
\eea

\bea 
M^{2\,(1)} & = &  
\bpm
\frac{\sqrt{2} r w \m_4}{1-r^2} & -\frac{(r^2+1) w \m_4}{\sqrt{2} (1-r^2)} & -\frac{r \m_4}{\sqrt{2}} & \k_1 ((\a_1 + r \a_2 + \a_4) - w^2 \rho_{12}) \\
 & \frac{\sqrt{2} r w \m_4}{1-r^2} & \frac{\m_4}{\sqrt{2}} & \k_1 (\a_2 + r (\a_1+\a_3)) \\ 
 & & 0 & w \k_1 (\rho_1 + \rho_2/2) \\
 & & & 0  
\epm\,\,\, , 
\label{Msq1}
\\
M^{2\,(0)} &= &
\bpm
\frac{r^2 \a_{34} + 2 w^2 \rho_{12}}{2(1-r^2)} &  \frac{r(\a_{34} + 2 w^2 \rho_{12})}{2(1-r^2)} & -w\rho_{12} & 0 \\ 
&   \frac{\a_{34} + 2 w^2 \rho_{12}}{2(1-r^2)} & 0 & 0 \\ 
& & \rho_{12} & 0 \\
& & & 2 \rho_1 
\epm \,\,\,\, ,
\label{Msq0}
\eea
where, for convenience, only the upper halves of the matrices $M^{2\,(1)}$ and
$M^{2\,(0)}$ are displayed in eqs.~\eqref{Msq1} and~\eqref{Msq0} respectively.

Since the smallest eigenvalue of $M^2$ must be of order $v^2$, its expression must
not contain any positive powers of $v_R$. Hence both $M^{2\,(0)}$ and $M^{2\,(1)}$
must have zero as their smallest eigenvalues. 
The other three eigenvalues of $M^2$ 
correspond to (mass)$^2$ of heavy CP-even scalars. This requires the non-zero eigenvalues
of $M^{2 \, (0)}$ to be positive.

We aim to calculate the smallest eigenvalue of $M^2$ using
perturbation theory. First we apply a similarity transformation on $M^{2}$, using the orthogonal matrix~\cite{Babu:2020bgz} 
which diagonalizes $M^{2\,(0)}$,
\bea 
O_I  = \bpm 
\frac{1}{k} & \frac{r}{k} & \frac{w}{k} & 0 \\
-\frac{r}{\sqrt{1+r^2}} & - \frac{1}{\sqrt{1+r^2}} & 0 & 0 \\
-\frac{ w}{k \sqrt{1+r^2}} &   -\frac{r w}{k \sqrt{1+r^2}} & -\frac{ \sqrt{1+r^2}}{k} & 0 \\
0  & 0 & 0 & 1
\epm \,\,\,\, ,
\eea 
where $k^2 = 1+r^2+w^2$. In the rotated basis the mass-squared matrix has the form
\begin{equation}
\begin{split}
\Tilde{M}^2=\Big(\Tilde{M}^{2\,(0)} v_R^2 + \Tilde{M}^{2\,(1)} v_R + \Tilde{M}^{2\,(2)}\Big)
     \end{split} \,\,.
\end{equation}
It is straight forward to check that both $\tilde{M}^{2\,(0)}$ and $\tilde{M}^{2\,(1)}$ 
have one zero eigenvalue. The positivity of the non-zero eigenvalues of $\tilde{M}^{2\,(0)}$
leads to the following constraints on the quartic couplings
\beq
2\rho_{12} = \rho_2 - 2 \rho_1 > 0~~{\rm and}~~ \a_{34} = \alpha_3- \alpha_4 > 0\,\,.
\label{qcineq}
\eeq

Using non-degenerate perturbation theory, the smallest eigenvalue of $M^2$, which we take to be  Higgs boson mass $m_h^2$, is found to be 
\bea
     m_h^2 & = & (\Tilde{M}^{2\,(2)})_{11}-\frac{[(\Tilde{M}^{2\,(1)})_{14}]^2}{2\rho_1}\,\,\,\nn\\
    \label{eq:e1}
     &=& \frac{\k_1^2}{2 (1+r^2 +w^2)}\times\nn\\
&&\Bigg( 4 \Big(\l_1 (r^2+1)^2 + 4 r (\l_4(r^2+1)+r \l_{23}) + w^2 (\a_{124} + r^2 (\a_1+\a_3) + \a_2 r) + \rho_1 w^4 \Big) \nn\\
&&\,\,\,\,\,\,\, -\frac{1}{\rho_1}(\a_{124} + r^2 (\a_1+\a_3) + \a_2 r + 2 \rho_1 w^2)^2 \Bigg),
\label{eq:mh}
\eea
where $\alpha_{124}=\alpha_1+r\alpha_2+\alpha_4$~\cite{Bernard:2020cyi}. Since $m_h^2$ should be positive, the second term in eq.~\eqref{eq:mh} must always be less than the first.
Thus the Higgs mass constraint imposes a rather strong restriction on the quartic couplings of DLRSM.

However, it has to be noted that for some specific set of values of quartic couplings, the 
assumptions of non-degenerate perturbation theory will not be valid. In these scenarios, the
smallest eigenvalue of $M^2$ matrix may differ significantly from the value given by the expression in eq.~\eqref{eq:mh}. We are interested in exploring the constraints on full parameter space of the quartic couplings. Therefore, we diagonalize the mass matrix numerically to avoid the pitfalls of  possible near degeneracies of the diagonal elements of $M^2$ matrix. As 
discussed in the 
introduction, we need a $v_R$ of about $20$ TeV to satisfy the experimental lower bound on
heavy charged gauge boson mass. Hence, we set $v_R = 20$\,TeV in our numerical work, which
also ensures that $\k_1^2, \k_2^2, v_L^2 \ll v_R^2$.

DLRSM contains the following extra scalars compared to the SM:
\begin{itemize}
    \item three CP-even scalars, labelled $H_{1,2,3}$ earlier in this section,
    \item two CP-odd scalars, labelled $A_1$ and $A_2$,
    \item two pairs of charged scalars, labelled 
    $H_1^\pm$ and $H_2^\pm$.
\end{itemize}
Among these, the CP-even scalar $H_3$ has some peculiar properties.
Its predominant component is the $SU(2)_R$ scalar $\chi_R^0$ and 
its mass is given by $m^2_{H_3} \approx 2 \rho_1 v_R^2$. Its couplings to SM fermions and
gauge bosons are highly suppressed. Hence mass constraints on neutral
scalars from neither flavour changing neutral interactions\,(FCNI) nor
direct searches are applicable to $H_3$. In principle, its mass
can be quite low~\cite{Dev:2016dja}, much lower than those of $H_1$ and 
$H_2$. The rest of the extra scalars can have significant coupling to
SM fermions and gauge bosons and hence all the experimental constraints
on heavy scalar masses will be applicable to them.

The gauge and the mass eigenstates of the neutral scalars are related by 
\bea 
X_\text{ph} = U^T X, \,\,\,\,\,\,\,\,\,\,\,\,\, U^T M^2 U = M^\text{2 diag}\,\,.
\label{eq:cp0scalardiag}
\eea
The coupling of scalars in the gauge basis to $W_1$ boson is 
\bea 
\mathcal{L} \supset \frac{1}{2} g^2 \, \big( (\k_1 - 2 c_{\xi} s_{\xi} \k_2) \phi^{0}_{1 r} + (\k_2 - 2 c_{\xi} s_{\xi} \k_1) \phi^{0}_{2 r} + v_L c_{\xi}^2 \chi_{L r}^{0} + v_R s_{\xi}^2 \chi_{R r}^0 \big)\,W^{+}_{1 \mu}\,W^{-\mu}_{1}\,\,\, , \nn \\
\eea
where $\xi$ is the mixing angle between $W_{L,R}$ given by eq.~\eqref{eq:weigenvec}.
Transforming the scalar gauge eigenstates to mass eigenstates, we find the 
coupling multiplier for the 3-point interaction of $h$ to $W_1$ bosons to be  
\bea 
\kappa_W = \frac{c_{h W_1 W_1}}{c_{hWW}^\text{SM}} =
\frac{1}{k} \Big((1 - 2 c_{\xi} s_{\xi} r) U_{11} + 
(r - 2 c_{\xi} s_{\xi}) U_{21}
+ w c_{\xi}^2 U_{31}
+ \frac{v_R}{\k_1} s_{\xi}^2 U_{41}
\Big),
\eea
where we expressed the DLRSM coupling in the form of a ratio to the corresponding SM coupling.
This makes is easier to compare our model calculations with the experimental results 
of Higgs boson couplings because they are given in the form of such ratios. Similarly, the coupling of scalars in gauge basis to the $Z_1$-boson is 
\bea 
\mathcal{L} \supset \frac{g^2}{4}\, (c_W + s_W s_Y)^2 (\k_1 \phi^{0}_{1 r} + \k_2 \phi^{0}_{2 r} + v_L \chi^{0}_{L r})\, Z_{1 \mu}\,Z^{\mu}_{1}\,\,\,,
\eea
which gives
\bea
\kappa_Z = \frac{g^2 (g^2 + 2 g_{BL}^2)}{(g^2+g'^2) (g^2 + g_{BL}^2)} \,\frac{1}{k} \,(U_{11}  + r U_{21} + w U_{31}) \,\,.
\label{eq:kappahzz}
\eea
Here $g'$ is the gauge coupling of $U(1)_Y$ of SM. We use $g_{BL} = g g'/(g^2 - g'^2)^{1/2}$ which reduces the fraction involving the gauge couplings in eq.~\eqref{eq:kappahzz} to unity~\cite{Georgi:1977wk}.

Triple Higgs~$(h^3)$ vertex in this model is given by
\bea 
 c_{h^3} &=& \frac{\k_1}{2} \Big(2 (\l_1 + r \l_4) U_{11}^3 + 2(r\l_1 + \l_4) U_{21}^3 + 2 w \rho_1 U_{31}^3 + 2(r(\l_1+4\l_2 + 2\l_3)+3\l_4) U_{11}^2 U_{21} \nn \\
 && \hspace{20pt} 
 +  2(\l_1+4\l_2 + 2\l_3 +3 \l_4 r) U_{11} U_{21}^2 + w (\a_1 + \a_4) U_{11}^2 U_{31}  + (\a_1 + r \a_2 + \a_4) U_{11} U_{31}^2 \nn\\
 &&  \hspace{20pt} 
 + w (\a_1 + \a_3) U_{21}^2 U_{31} + (\a_2 + r(\a_1 + \a_3)) U_{21} U_{31}^2   \Big) \,\,\, ,
 \label{kappah}
\eea
with the corresponding coupling multiplier being $\kappa_h = c_{h^3}/c_{h^3}^\textrm{SM}$ where $c_{h^3}^\textrm{SM} = m_h^2/2 v$.

\subsection{The fermion sector}
\label{sec:fermionsector}

The fermions obtain masses through their Yukawa couplings to the bidoublet. For the quark sector, these couplings are
\bea 
\mathcal{L}_{Yuk} \supset - \bar{Q}_{Li} (y_{ij} \Phi + \tilde{y}_{ij} \tilde{\Phi}) Q_{Rj} + h.c. \,\, , 
\label{eq:uvyukawaterms}
\eea
which lead to the following mass terms
\bea 
\mathcal{L}_{m} \supset - \bar{U}'_{Li} (M_U)_{ij} U'_{Rj} -\bar{D}'_{Li} (M_D)_{ij} D'_{Rj} + h.c. \,\,.
\eea
Here $U' \equiv (u', c', t')$ and $D' \equiv (d', s', b')$ are up- and down-type quarks in the gauge basis. The mass matrices $M_{U (D)}$, in general, are complex and arbitrary
\bea
M_U = \frac{1}{\sqrt{2}}(\k_1 y + \k_2 \tilde{y}), \,\,\,\,
M_D = \frac{1}{\sqrt{2}}(\k_2 y + \k_1 \tilde{y})\,\,.   \nn
\eea
They must be diagonalized to identify the quark mass eigenstates. 
Interactions of the gauge eigenstates of the neutral CP-even scalars to quarks are given by~\cite{Deshpande:1990ip}
\bea 
\mathcal{L}_N &\supset &  \frac{\k_1}{\sqrt{2}\,\k_{-}^2}\, \bar{u}_{Li} \Bigg((\phi^{0}_{1} - r \phi^{0*}_2) \hat{M}_u + (-r \phi^{0}_{1} + \phi_2^{0 *}) V_{L}^{\text{CKM}}\,\hat{M}_d\,V_{R}^{\text{CKM} \dagger}   \Bigg) u_{Rj}  \nn\\
&& + \frac{\k_1}{\sqrt{2}\,\k_{-}^2}\, \bar{d}_{Li} \Bigg((\phi^{0*}_{1} - r \phi^{0}_2) \hat{M}_d + (-r \phi^{0*}_{1} + \phi_2^0) V_{L}^{\text{CKM} \dagger}\,\hat{M}_u\,V_{R}^{\text{CKM}}   \Bigg) d_{Rj} + h.c.    \,\,,
\label{eq:hffcoupling}
\eea
where $\k_-^2 = \k_1^2 - \k_2^2 = (1 - r^2) \k_1^2$ and $\hat{M}_{u (d)}$ is the diagonal up~(down)-type quark matrix\footnote{In the gauge basis of the scalars, the Yukawa sector of DLRSM is the same as a triplet left-right symmetric model of ref.~\cite{Deshpande:1990ip}.}. 
As we consider a scalar sector that conserves CP at the tree-level, manifest left-right symmetry holds in the quark sector,\ie $V_{R}^{\text{CKM} } = V_{L}^{\text{CKM}}$. We use the Wolfenstein parametrization of the CKM matrix with best-fit values of relevant parameters~\cite{pdg:ckm}
\bea 
A = 0.79,\,\,\, \l = 0.2265,\,\,\, \bar{\rho} = 0.141,\,\,\, \bar{\eta} = 0.357   \,\,  .  \nn
\eea 
Using eqs.~\eqref{eq:cp0scalardiag}  and~\eqref{eq:hffcoupling} we obtain the couplings of CP-even scalars with quarks. 
We find that the couplings of fermions to the lightest scalar $h$ are essentially flavour diagonal with small flavour violating couplings of
the order of $v^2/v_R^2$. The heavier scalars, with masses of the order of $v_R$, have flavour violating couplings in general. For $v_R = 20$ TeV, the amplitudes of FCNI are below the experimental upper bounds. The coupling of the CP-even lightest scalar to third generation quarks are given as
\bea 
c_{htt\,(hbb)} = \frac{\k_1}{\sqrt{2}\,\k_{-}^2}\,\Big((U_{11}-r U_{21})m_{t(b)} + (U_{21}-r U_{11}) (V_{L}^{\text{CKM}}\,\hat{M}_{d(u)}\,V_{R}^{\text{CKM} \dagger})_{33} \Big), 
\label{eq:htthbb}
\eea 
which can be used to obtain the coupling multipliers $\k_t$ and $\k_b$, as $\kappa_{f} = c_{hff}/c_{hff}^\text{SM}$, where $c_{hff}^\text{SM} = m_f/v$ and $f=t$ or $b$.

\section{Theoretical constraints}
\label{sec:theoreticalbounds}

Here we briefly discuss the theoretical constraints on DLRSM, such as perturbativity of the quartic and Yukawa couplings, perturbative unitarity of gauge boson scattering, and boundedness of the scalar potential from below. 

\subsection{Perturbativity of the quartic and Yukawa couplings}

Perturbativity bounds on the quadratic couplings dictate $\a_i, \l_i, \rho_j \lesssim 4 \pi$, where $i = 1, 2, 3, 4$ and $j = 1,2$. Moreover, we also assume $\mu_4 \lesssim v_R$. 

\begin{figure}[h!]
 \begin{center}
 \includegraphics[width=4.in,height=3.7in, angle=0]{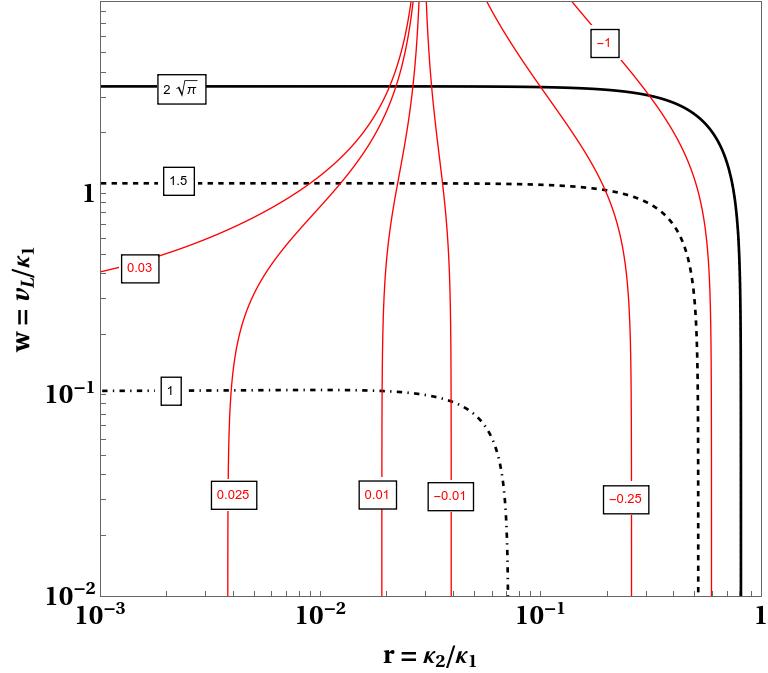} 
\caption{The black and red contours represent the values of $y_{33}$ and $\tilde{y}_{33}$ respectively.  }
\label{fig:ytperturbativeexclusion}
\end{center}
\end{figure}

Perturbativity of the Yukawa couplings $y$ and $\tilde{y}$ appearing in eq.~\eqref{eq:uvyukawaterms} pose a strong bound on the $r-w$ plane. In the limit $\Big(V_{L,R}^\text{CKM}\Big)_{33} \sim 1$, 
\bea 
y_{33} = \frac{\sqrt{2} (1+r^2 +w^2)^{1/2}}{v (1 - r^2)} \,(m_t - r \,m_b)\,\,\,, \nn\\
\tilde{y}_{33} = \frac{\sqrt{2} (1+r^2 +w^2)^{1/2}}{v (1 - r^2)} \,(m_b - r \,m_t)\,\,\,.
\eea   
From these equations, we note that $y_{33} \simeq 1$ and $\tilde{y}_{33} \simeq 1/40$
for very small values of $r$ and $w$. 
The perturbativity requirement of $y_{33} < (4 \pi)^{1/2}$ leads to strong upper 
limits $r \lesssim 0.8$,  and $w \lesssim 3.5$. The perturbativity requirement of $\tilde{y}_{33}$ gives comparatively weaker constraints. These constraints are shown in fig.~\ref{fig:ytperturbativeexclusion}.

\subsection{Perturbative unitarity of gauge boson scattering}

The condition of unitarity of the $2 \to 2$ scattering of the gauge degrees of freedom impose certain constraints on the masses and couplings of the new scalars. Some of these constraints are superseded by perturbativity for $r < 1$. But the following conditions must be satisfied at $\mathcal{O}(\k_1/v_R)$~\cite{Bernard:2020cyi}
\bea
0 <   \rho_1  <  \frac{8 \pi}{3}\, , \,\,\text{or,}\,\, \frac{m_{H_3}^2}{v_R^2} & < & \frac{16 \pi}{3} \,\,\,,  \nn\\ 
\frac{(c_{H_3})^2}{k^4} \, \frac{m^2_{H_3}}{v_R^2} &<& \frac{16 \pi}{3} \,\,\, , \nn\\
2 \frac{w^2}{k^2} \sum_{i=1,2} F_i^2 \frac{m^2_{H^\pm_i}}{v_R^2} + \frac{c_{H_3}}{k^2} \frac{m^2_{H_3}}{v_R^2} & < & 16 \pi\,\,\,, \nn\\
2 \frac{w^2}{k^2} \sum_{i=1,2} S_i^2 \frac{m^2_{H^\pm_i}}{v_R^2} + \frac{c_{H_3}}{k^2} \frac{m^2_{H_3}}{v_R^2} & < & 16 \pi\,\,\, ,
\label{eq:unitaritycondition}
\eea
where the expressions for $c_{H_3}$, $S_i$ and $F_i$, in terms of the parameters 
of the Higgs potential, are given in ref.~\cite{Bernard:2020cyi}. 

\subsection{Boundedness from below}

Criteria for boundedness from below of the DLRSM scalar potential has been studied in the literature utilising the copositivity criteria~\cite{Chakrabortty:2013zja,Chakrabortty:2016wkl} and gauge orbit formulation~\cite{Chauhan:2019fji,Frank:2021ekj,Kannike:2021fth}. These studies used different
simplified forms of the potential. We have derived the criteria for boundedness from 
below for the most general potential given by eq.~\eqref{eq:scalarpotential}. These are
described in Appendix~\ref{appendix:stability}.

\subsection{Custodial symmetry breaking}
\label{sec:ewpt}

Some key constraints on the extended scalar sectors arise from the measurement of the $\rho$ parameter, which quantifies the breaking of the custodial symmetry of the scalar potential in the SM. The tree-level contribution to $\rho = m_{W_1}^2/m_{Z_1}^2 \cos \theta_W^2$ in the DLRSM appears at the order $(v/v_R)^2$~\cite{Bernard:2020cyi}.

\section{A convenient parametrization of the quartic couplings}
\label{sec:convenient}
 
The model contains ten quartic couplings overall, four $\l$s, four $\a$s and
two $\rho$s. These couplings have to satisfy various inequalities arising from
the theoretical constraints discussed in the previous section and the positivity 
of CP-even scalar masses. There is an inequality on the ratio 
$x=\l_2/\l_4$ arising from boundedness of the 
scalar potential. As shown in Appendix~\ref{appendix:stability}, this inequality
is of the form $8x (1 - x) > 0$. For positive $x$, this translates into
$0.25 \leq x \leq 0.85$.

We are interested in how the Higgs data constrain ranges of $r$ and $w$.
In studying these constraints, in principle, the ten quartic couplings 
should be allowed to take ten independent values,
subject to the inequalities mentioned above. However, a study with ten
free parameters does not give a good insight into various inter-relations.
To gain an insight as to how the data constrains $r$ and $w$, we 
pick a simplified set of three quartic couplings $\l_0, \a_0$ and $\rho_1$. We fix the ten quartic couplings by using relations
\bea 
 \{ \l_1 = \l_3 = \l_4 = \l_0,\,\,\, x = \frac{\l_2}{\l_4},\,\,\, \a_1 = \a_2 = \a_4 = \a_0, \, \, \, p = \frac{\a_3}{\a_4} - 1,\,\,\,  q = \frac{\rho_2}{2 \rho_1} - 1 \}.
\eea
The set of parameters $\{\l_0, \a_0, \rho_1, x, p, q, r, w, v_R \}$ together with $v$
provides the most minimal simplified basis to describe the scalar potential of the model.
From now on we refer to this as the `{\it simple}' basis, as opposed to the `{\it generic}' basis where all the $\l$s and all the $\a$s are taken to be independent. We
note that, both in the simple basis and in the generic basis, $p$ and $q$ parameters need to be positive.

\section{Constraints from the Higgs data}
\label{sec:higgsbounds}

In sec.~\ref{sec:theoreticalbounds}, we derived the upper bounds on $r$ and $w$
from theoretical constraints. In this section, we investigate the constraints on these parameters coming from the Higgs data. We first do
this study in the {\it simple} basis of the DLRSM quartic couplings
to see the pattern of allowed regions in the $r-w$ plane. We then 
re-do the study in the {\it generic} basis to see what changes, if any,
are there in our results. 

\subsection{Constraints in the \textit{simple} basis} 
\label{sec:simplebasis}

In the {\it simple} basis, there are eight parameters,
three quartic couplings, the three ratios $p$, $q$ and $x$, the coupling of
three scalar multiplets $\mu_4$, and $v_R$. In our calculations, we fix the 
values of $v_R = 20$ TeV. We vary the other parameters 
randomly and identify regions in the parameter space which satisfy the 
experimental constraints from the Higgs data.

The mass of the lightest Higgs scalar depends strongly on the quartic couplings $\l$, as shown in eq.\,\eqref{eq:mh}. Hence, the constraint on $m_h$ will be satisfied only for a limited range of values of $\l_0$. Therefore, in choosing random values of this parameter, we limit ourselves to a restricted range. For specific values of  $\a_0$, $\rho_1$, $x$, $p$, and $q$, we solve eq.\,\eqref{eq:mh} and obtain a solution for $\l_0$ which we label $\Lambda_0$. To maximize the number of points satisfying the $m_h$ constraint, we choose random values of $\l_0$ in the limited range 
\bea 
\l_0 = (1 + y) \, \Lambda_0\,\, {\rm with} \, \, y \in [-0.1,0.1].
\eea
This choice ensures that most of the random points satisfy the Higgs mass constraint and enables us to study how the constraints from the $\k$ factors restrict the parameters of the model. 

The random values of $x$ and $y$ are picked in linear scale in their
respective ranges. For the other parameters, the random values are picked in a log scale, in the ranges specified below:  
\begin{equation}
    \a_{1,2,4} \equiv \a_0 \in [10^{-3}, 4 \pi], \, \, 
    \rho_1 \in [0.1, 8\pi/3], \, \, 
    \mu_4 \in [10^{-2}, 1]\times v_R.
\label{eq:scanrange1}
\end{equation}
This choice of using logarithmic scale, leads to very small values of the parameters being as likely as the values of the order $1$. Of the two ratios,
$p$ and $q$, we hold $q$ fixed at $q=1$ but consider four different values for $p$, varying from a low value of $0.02$ to a high value of $5$.

\begin{table}[]
    \centering
\begin{tabular}{c|c}
 Observable & Observed value   \\
\hline
$m_h$ & $(125.38 \pm 0.14)$\,GeV~\cite{CMS:2020xrn}  \\
$\k_W$ &   $1.05 \pm 0.06$~\cite{ATLAS:2020qdt}\\
$\k_Z$ & $0.96\pm 0.07$~\cite{CMS:2020gsy} \\
$\k_h$ & $[-2.3, 10.3]$ at $95\%$\,CL~\cite{ATLAS:2019pbo}   \\
$\k_t $ & $1.01\pm 0.11$~\cite{CMS:2020gsy}\\
$\k_b$ &   $0.98^{+0.14}_{-0.13}$~\cite{ATLAS:2020qdt}\\
\hline
\end{tabular}
\caption{Experimental values of the Higgs mass and coupling multipliers used in our calculations. }
\label{table:observed values}
\end{table}

\begin{figure}[h!]
\centering
\subfigure[]{
\includegraphics[width=3in,height=2.85
in, angle=0]{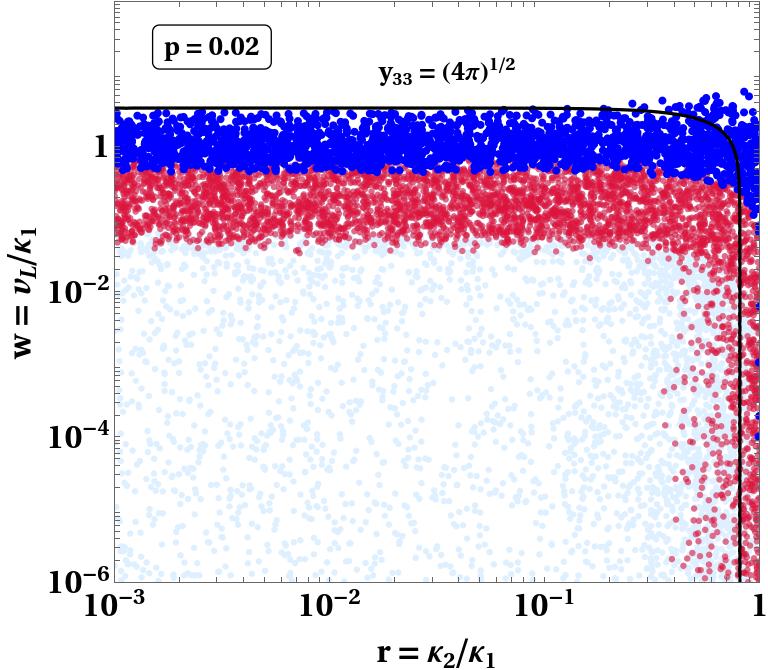} }
\hskip 5pt
\subfigure[]{
\includegraphics[width=3in,height=3in, angle=0]{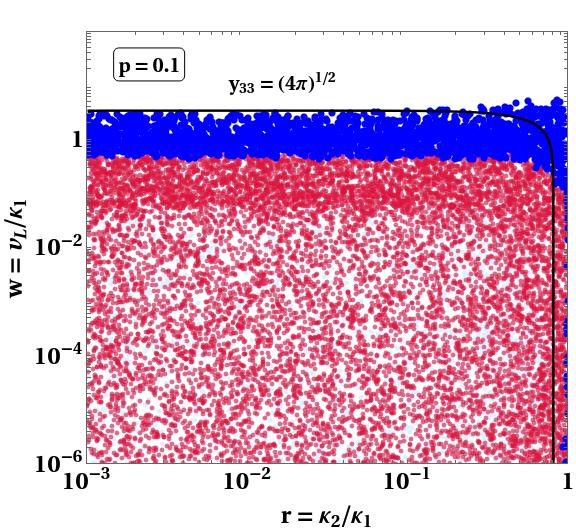} }
\subfigure[]{
\includegraphics[width=3in,height=2.85in, angle=0]{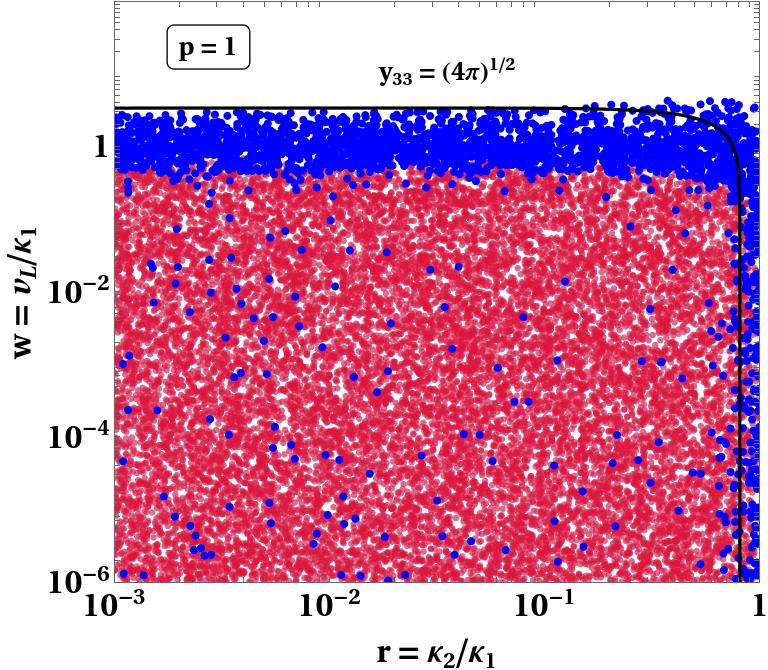}  }
\hskip 15pt
\subfigure[]{
\includegraphics[width=3in,height=3in, angle=0]{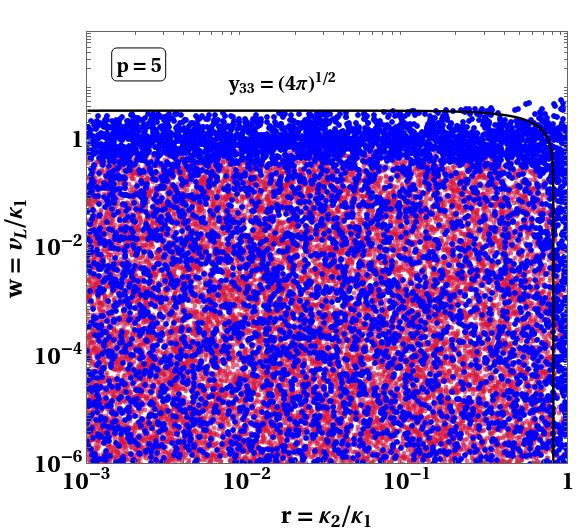}  }
\caption{Allowed points on the $r$-$w$ plane  with random values of $\a_{1,2,4} = \a_0 \in [10^{-3}, 4\pi]$, $\l_{1,3,4} = \l_0 \in [10^{-3}, 4\pi]$, $x = \l_2/\l_0 \in [0.25, 0.85]$, $\rho_1 \in [0.1, 4\pi]$ for four
different values of $p$. The values of $q$ and $v_R$ are held fixed at
$q=1$ and $v_R=20$\,TeV respectively. Light blue points satisfy  $m_h$, $\k_h$,  $\k_W$, $\k_Z$ and $\k_t$ constraints and theoretical bounds. Red points satisfy $\k_b$ constraint also and dark blue points satisfy the additional constraint $m_{H_1} > 15$ TeV.} 
\label{fig:rwsimple}
\end{figure}

For each random point, we numerically calculate the lightest Higgs mass $m_h$ and the coupling parameters $\k_W$, $\k_Z$, $\k_h$, $\k_t$ and $\k_b$ and demand that they should match the corresponding experimentally measured values, listed in Table~\ref{table:observed values}. The results of our calculation are displayed in fig.~\ref{fig:rwsimple}. The light blue dots satisfy the theoretical bounds and the constraints from $m_h$, $\k_h$, $\k_W$, $\k_Z$ and $\k_t$. By considering these constraints individually, it was found
the $\k_h$ constraint rules out a significant number of points allowed by the $m_h$ constraint. The $\k_W$, $\k_Z$ and $\k_t$ constraints are satisfied by all the points which satisfy $m_h$ and $\k_h$ constraints. The red dots satisfy the additional  constraint from $\k_b$ parameter. The $\k_b$ constraint leads to a drastic reduction in the number of allowed points for low values of $p$. In particular, for $p=0.02$, very low values of $r$ and $w$ are ruled out. As the value of $p$ increases, points of low $r$ and low $w$ become allowed but there is a clear preference for larger values of $w$ as long as $p \leq 0.1$. Only for $p \geq 1$, the distribution of allowed points is uniform over the full range of $r$ and $w$. This means we need to have $\a_3$ at least twice as large as the other $\a$\,s, if the usual assumption $v_L, \k_2 \ll \k_1$ is to hold. 

The reason for this strong influence of $\k_b$ can be understood from eqs.~\eqref{Msq0} and \eqref{eq:htthbb}. In the expression for $\k_b$
in eq.~\eqref{eq:htthbb}, the second term is proportional to $m_t$. If the CP-even scalar mixing matrix element $U_{21}$ is not negligible, this second term can lead to a large value of $\k_b$. The dominant contribution to 
$m_{H_1}^2$ comes from $v_R^2 M^{2 (0)}_{22} = v_R^2 \,[p\, \a_3 + 2 w^2 \rho_{12}]/[2(1 - r^2)]$ in eq.~\eqref{Msq0}. If  $M^{2 (0)}_{22}$ is too small, there will be a significant mixing between $h$ and $H_1$, \ie  $U_{21}$ is not negligible and the predicted $\k_b$ will deviate significantly from $1$. 
A large value for $M^{2 (0)}_{22}$ can be obtained for $r \simeq 1$, independent of the values of  $p$ and $w$, but this violates the perturbativity of the Yukawa coupling $y_{33}$ and is not allowed. For moderate values of $r$, either $p$ or $w$ (or both) must take reasonably large values to satisfy the constraint on $\k_b$. We see this feature
illustrated in fig.~\ref{fig:rwsimple} where small values of $w$ are ruled out by the $\k_b$ constraint when $p$ is small but become progressively more allowed as $p$ becomes larger. 

In addition to the constraint from the measured parameters listed in Table~\ref{table:observed values}, there is an important indirect bound on the masses of heavy CP-even scalars from flavour changing neutral interactions. The requirement that these interactions be adequately suppressed leads to a strong bound~\cite{Zhang:2007da}
\bea
    m_{H_1} \geq 15\,{\rm TeV}.
\eea
In fig.~\ref{fig:rwsimple}, dark blue dots represent the points which satisfy this additional constraint also. For $p \leq 1$, the blue
dots are bunched towards values of $w \geq 0.6$. The few points of low $w$, which occur for $r \simeq 1$ are ruled out by the constraint $r \leq 0.8$, arising from the perturbativity bound on the Yukawa coupling $y_{33}$. 
Therefore, the bound on $m_{H_1}$ strongly prefers that \textit{vev} of $\chi_L$ has a significant influence on the breaking of $SU(2)_L$. For very large values of $p \geq 5$, we see a reasonably large number of blue dots in the region of low $r$ and low $w$. 
For example, in fig.~\ref{fig:rwsimple}\,(d), the range $r \in [10^{-3}, 10^{-2}]$ and $w \in [10^{-6}, 10^{-3}]$ contains $\sim 14\%$ of the total number of points allowed from the measurements of $m_h$, $\k_h$, $\k_t$, $\k_b$, and $m_{H_1} > 15$\,TeV. These points correspond to larger values of $\rho_1 \gtrsim 1$ which push $\rho_2 > 4$ 
making it relatively large.
That is, the usual assumption $v_L, \k_2 \ll \k_1$ holds, only if the quartic couplings $\rho_1$ and $\rho_2$ take moderately large values.

In the expression for $\k_h$ in eq.~\eqref{kappah}, the terms proportional to $w$ always make a positive contribution to the value of $\k_h$. Hence, the experimental upper limit on $\k_h$ affects the region of large $w$ significantly. 
In future, if this upper limit is improved to 
$\k_h < 2$, 
values of $w > 1$ will be ruled out. Hence the points satisfying all the constraints (dark blue points of fig.~\ref{fig:rwsimple}) will occur only for very large 
values of $p \sim 5$ with allowed values of $w < 0.5$.

We now briefly discuss what happens when the ratio $q$ is allowed to deviate from $1$. From eq.~\eqref{Msq0}, we see that leading contribution to the mass of one of the heavier Higgses is $\rho_{12} v_R^2 = \rho_1 q v_R^2$. Hence, very small values of $q$ will not satisfy $m_{H_1} > 15$ TeV constraint whereas larger values of $q$ will violate unitarity bounds. These two constraints together restrict
$q$ to the range $0.01 \leq q \leq 4$.  The constraints from $m_h$, $\k_h$, $\k_W$, $\k_Z$, $\k_t$ and $\k_b$ are unaffected by changes in the value of $q$. 
As a result, the points satisfying all the constraints (dark blue points of fig.~\eqref{fig:rwsimple}) are depleted overall if $q$ deviates from 1. 
For the extreme allowed values of $q$,\ie $q \sim 0.01$ and $q \sim 4$, $m_{H_1} > 15$ TeV is more likely to be satisfied for $w \gtrsim 1$. As mentioned above, these extreme values will be ruled out if $\k_h$ is measured to be less than $2$.

In the analysis above, the constraints on $m_h$ and $m_{H_1}$ were discussed thoroughly. Here we briefly comment on the masses of the additional scalars in the model.
The algebraic expressions for their masses are given in Appendix A. We find that $m_{H_1}, m_{A_1}$ and $m_{H_1^\pm}$ are all highly degenerate with their masses differing by less than $1\%$, for the values of quartic couplings which satisfy all the experimental constraint listed in Table~\ref{table:observed values}. We also find that the masses of other scalars, $m_{H_2}, m_{A_2}$ and $m_{H_2}^\pm$ are similarly degenerate but at a higher values compared to $m_{H_1}$.

\subsection{A strong upper limit on $v_L$ from Higgs mass}

\begin{figure}[htb]
\centering
 \includegraphics[width=3.2in,height=3.in, angle=0]{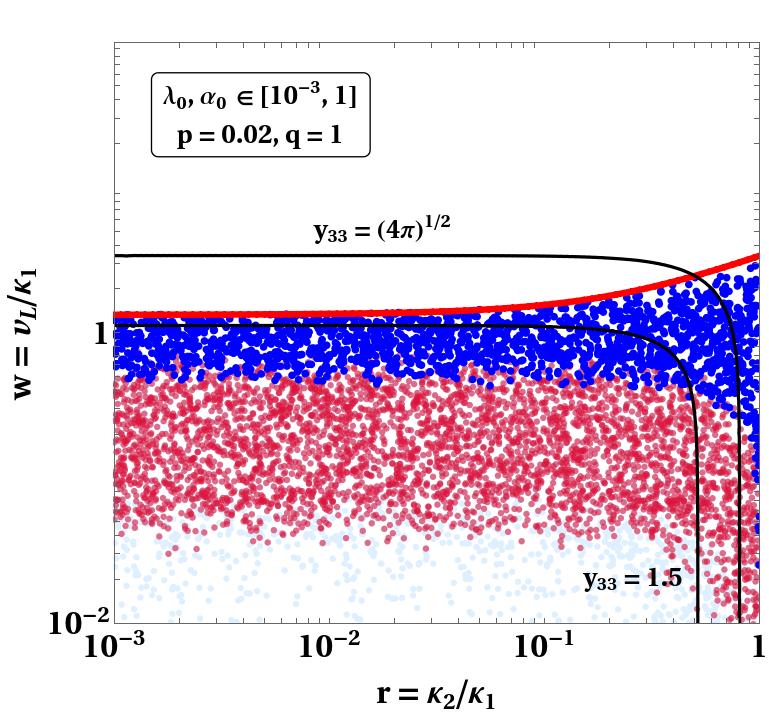}  
\caption{Allowed points on the $r$-$w$ plane for $\l_i , \a_i \in [10^{-3}, 1] $, $\rho_{1} \in [0.1, 8 \pi/3]$, and $\rho_2 \in [0.1, 4 \pi]$. The light blue
points satisfy $m_h$, $\k_W$, $\k_Z$, $\k_h$ and $\k_t$ constraints and theoretical bounds. The red
points satisfy $\k_b$ constraint in addition and the dark blue points satisfy
$m_{H_1} > 15$ TeV constraint. All the allowed points are below the red 
line, which represents the analytical upper limit of $w$.  The black solid lines represent contours of Yukawa couplings $y_{33}$.}
\label{fig:analyticalWmax}
\end{figure} 

As mentioned earlier, the parameter space with $w \gtrsim 1$ is of special interest. It was seen in fig.~\ref{fig:rwsimple} that there is a strong upper 
bound on $w$ arising from the measurement of $m_h$. In the \textit{simple} basis eq.~\eqref{eq:mh} takes the form
\bea 
m_h^2 = \frac{v^2}{(1+r^2+w^2)^2} \Big[ 2\big(8 \l_2 r^2 + \l_0 (1+r^2)^4\big) - \frac{\a_0^2 (2 + r(2 + (2+p)r) )^2}{2 \rho_1} \Big]\,\,.
\label{eq:mhsimple}
\eea
In the limit $\a_0 \rightarrow 0$ and $r \rightarrow 1$, imposing the perturbativity bounds $\l_0, \l_2 < 4 \pi$ leads to the upper limit $w < 6.81$. Demand of boundedness from below requires $\l_2 < 0.85\,\l_0$, which in turn, slightly reduces the upper limit to $w < 6.71$. The upper limit can also be obtained as a function of $r$, $w_\text{max} (r) \simeq a + b r + c r^2 $, where $a = 2.9332$, $b = 4.3535$, and $c = -0.4755$. 
If the maximum values of the couplings $\a_0$ and $\l_{0,2}$ are taken to be $1$ instead of $4\pi$, the upper limit reduces to $w_\text{max} \sim 2.8$ at $r \sim 1$. In fig.~\ref{fig:analyticalWmax} we have illustrated this analytical bound on top of a random scan over $\a_0, \l_0 \in [10^{-3},1]$ which shows that all the allowed points are below the limit imposed by the Higgs mass constraint of eq.~\eqref{eq:mhsimple}. Note that, the perturbativity limit on $y_{33}$ is stronger than the Higgs data for $r \gtrsim 0.5$. As a result, the effective upper limit on $w$ as a function of $r$ is set by the interplay of $m_h$ and perturbativity of $y_{33}$.

\subsection{Constraints in the \textit{generic} basis}
\label{sec:genericbasis}

In the {\it generic} basis, each quartic coupling is allowed to take a
different value. The total number of parameters in this case is eleven: 
ten quartic couplings (four $\a$\,s, four $\l$\,s and two $\rho$\,s) and 
$\mu_4$. As in the {\it simple} case, we parametrize $\a_3 = \a_4 (1+p)$ and
$\rho_2 = 2 \rho_1 (1 + q)$. Random values of these quartic couplings 
are varied over the ranges
\begin{equation}
    \l_2, \l_3, \l_4  \in [10^{-3}, 4 \pi], \,\,
    \a_1, \a_2, \a_4 \in  [10^{-3}, 4 \pi], \,\,
    \rho_1 \in [0.1, 8 \pi/3].
    \label{eq:scanrange2}
\end{equation}
As before, these random values are picked in a log scale.  
As in the {\it simple} case, to maximize the number of points satisfying the $m_h$ constraint we choose random values of $\l_1$ in the limited range 
\bea 
\l_1 = (1 + y) \, \Lambda_1\,\, {\rm with} \, \, y \in [-0.1,0.1],
\eea
where $\Lambda_1$ is the solution for $\l_1$ obtained by solving eq.~\eqref{eq:mh} for a specific set of values of $\l_{2,3,4}$, $\alpha_{1,2,4}$, $\rho_{1}$, and $p$. The random values of $y$ were chosen on a linear scale.
The numerical calculations here also are done for $v_R = 20$\,TeV and for four different values of $p$ varying from $0.02$ to $5$. The value of $q$, as before, is fixed at $q=1$.

\begin{figure}[h!]
\centering
\subfigure[]{\includegraphics[width=3.in,height=3.in, angle=0]{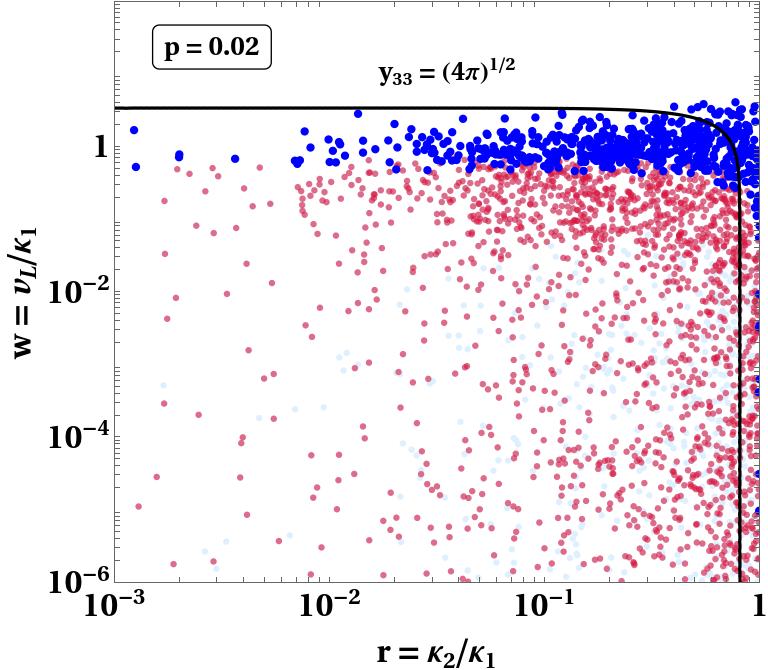} }
\hskip 10pt
\subfigure[]{\includegraphics[width=3.in,height=3.in, angle=0]{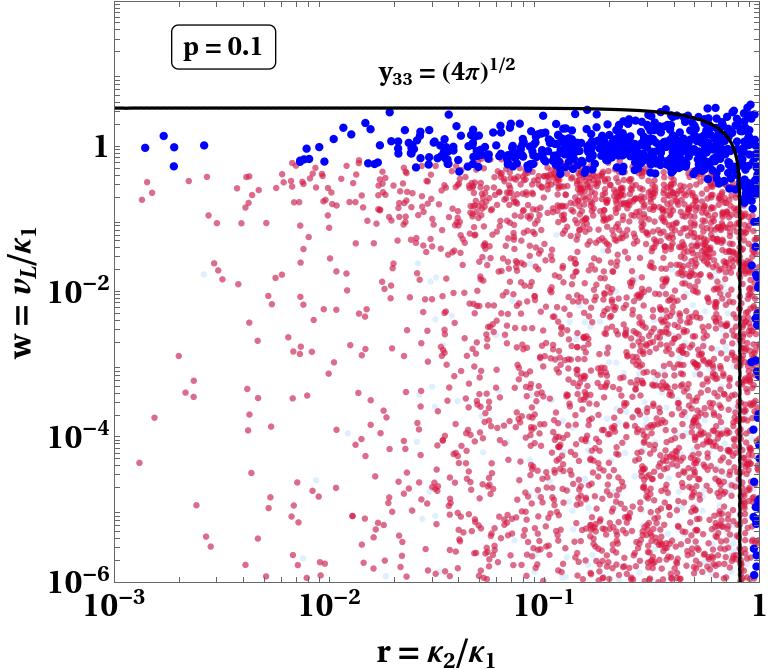} }
\subfigure[]{\includegraphics[width=3.in,height=3.in, angle=0]{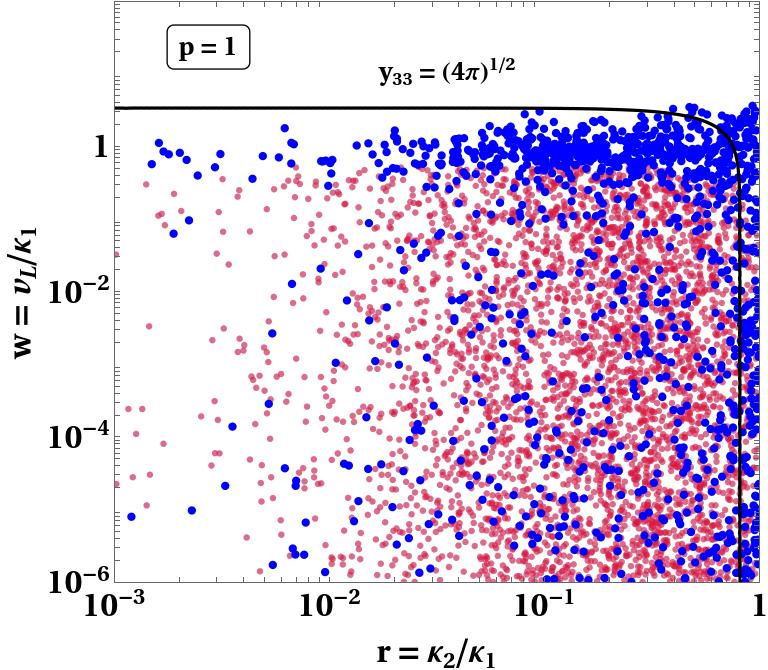} }
\hskip 10pt
\subfigure[]{\includegraphics[width=3.in,height=3.in, angle=0]{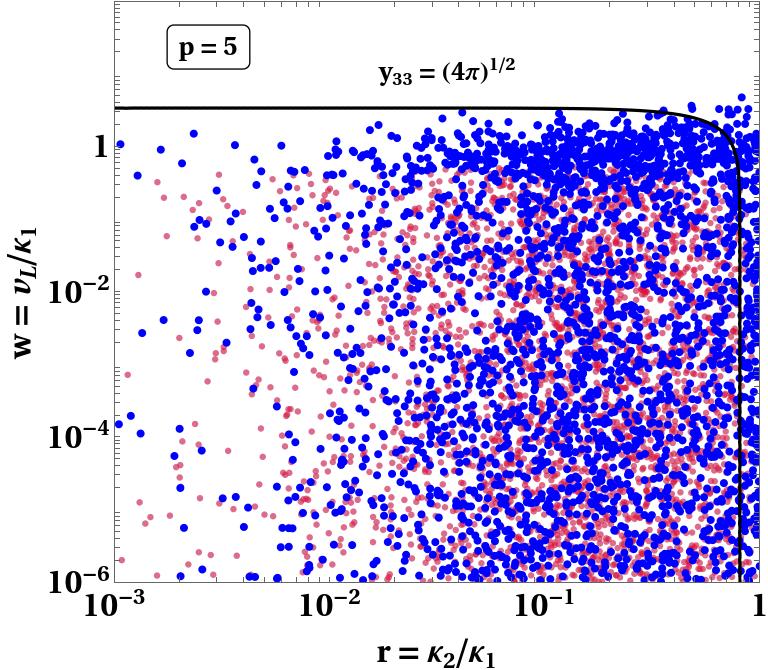} }
\caption{Allowed points on the $r$-$w$ plane  with random values of $\a_{1,2,4} \in [10^{-3}, 4\pi]$, $\l_{2,3,4} \in [10^{-3}, 4\pi]$, $\rho_1 \in [0.1, 8\pi/3]$, $q = 1$, while (a) $p = 0.02$, (b) $p = 0.1$, (c) $p = 1$, and (d) $p=5$. The colour coding is the same as fig.~\ref{fig:rwsimple}.
}   
\label{fig:rwgeneric}
\end{figure}

The result of this scan can be seen in fig.~\ref{fig:rwgeneric} for
the four different values of $p$. As in the case of the {\it simple} basis,
we have
\begin{itemize}
    \item the light blue points satisfying the theoretical bounds and constraints from $m_h$, 
    $\k_h$, $\k_W$, $\k_Z$, and $\k_t$,
    \item the red points satisfying the additional constraint from $\k_b$,
    \item the dark blue points satisfying the constraint from the lower limit on the heavy Higgs mass.
\end{itemize}

We note that the patterns of allowed points as a function of $w$, 
observed in fig.~\ref{fig:rwsimple} are repeated in fig.~\ref{fig:rwgeneric}. One contrast is that
fig.~\ref{fig:rwgeneric} contains very few allowed points at low values of $r$ compared to fig.~\ref{fig:rwsimple}. That is, if the quartic couplings are allowed to vary randomly without any pattern, it is more unlikely that 
$r$ and $w$ take very small values. Hence, for a generic set of values of
quartic couplings in DLRSM, the probability that $SU(2)_L$ breaking occurs essentially due to the \textit{vev} of $\phi_1^0$ is very small. If such a feature is desired, then the quartic couplings should be fine-tuned to particular values.

\subsection{Decoupling leads to Alignment}

\begin{figure}[htb]
\centering
\subfigure[]{
\includegraphics[width=2.9in,height=3.in, angle=0]{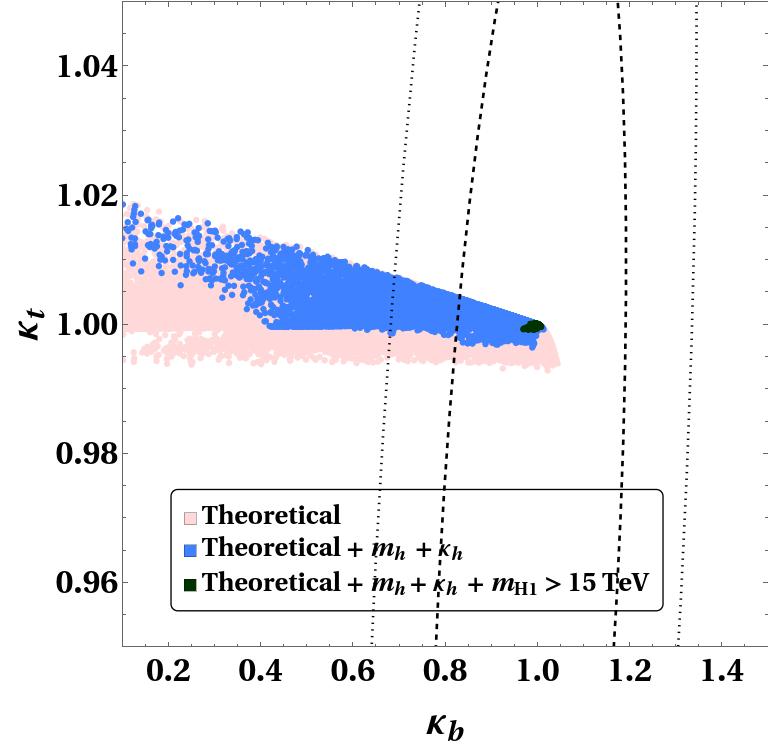}    }
\hspace{5pt}
\subfigure[]{
 \includegraphics[width=3.in,height=3.in, angle=0]{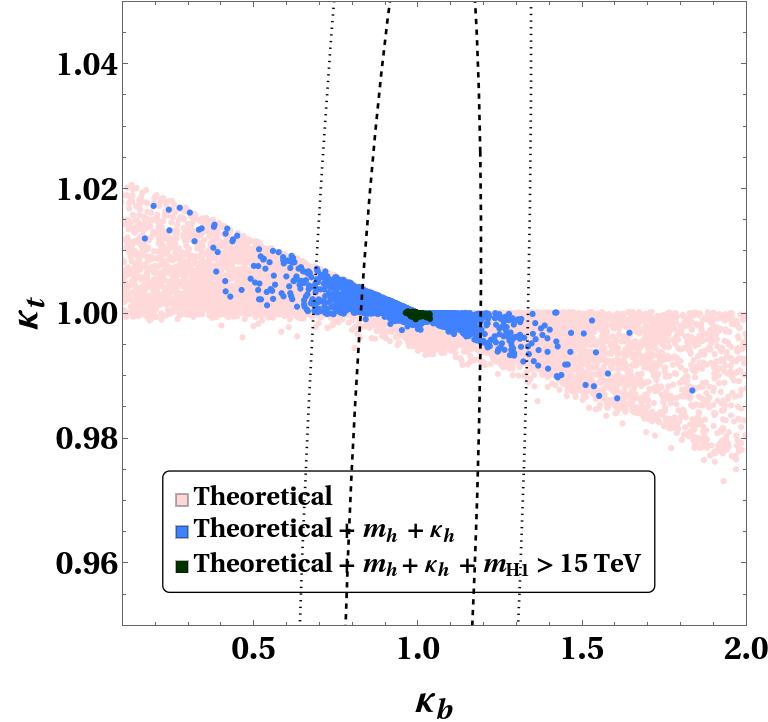}}
\caption{Predictions for $\k_b$ and $\k_t$ for (a) {\it simple} and (b) {\it generic} basis with random values of quartic couplings in the ranges given by eqs.~\eqref{eq:scanrange1} and ~\eqref{eq:scanrange2} respectively.  Additionally, $p$ and $q$ are varied logarithmically in the range $[0.1, 10]$. The pink points satisfy the theoretical constraints, the blue points satisfy the additional $m_h$ and $\k_h$ constraints, and dark green points also satisfy constraint on heavy Higgs mass $m_{H1} > 15$~TeV.}   
\label{fig:kbktAllowed}
\end{figure}

In this subsection, we discuss the interplay of various constraints from the Higgs data on the allowed values of parameters. All the parameters
of the model, including $p$ and $q$, are varied in the ranges described 
previously in sec.\,\ref{sec:simplebasis} and in sec.\,\ref{sec:genericbasis}.
We have calculated the values of $\k_b$ and $\k_t$ for each of these points 
and plotted them in fig.~\ref{fig:kbktAllowed} in the 
$\k_b-\k_t$ plane.

The left panel of fig.~\ref{fig:kbktAllowed} is for the \textit{simple} basis and the right panel is for the \textit{generic}
basis. The pink points satisfy the theoretical constraints. We note that all pink points satisfy the experimental constraint
on $\k_t$ but only a fraction of them satisfy the constraint on
$\k_b$. The points which satisfy the constraints on $m_h$, $\k_W$, $\k_Z$, and
$\k_h$ form a small subset of pink points and are highlighted as blue points. 
We see that the $\k_b$ constraint disallows about $9\%$ blue points. Finally,
we impose the constraint on heavy Higgs mass $m_{H_1} > 15$ TeV. The points
which satisfy this last constraint are shown in 
dark green. These green points
essentially form a very small fraction ($2.5\%$ of pink points) which are clustered around the point
$\k_b =1$ and $\k_t=1$. This clustering around SM values of $\k_b$ and $\k_t$
is analogous to the scenario of `alignment by decoupling' in two-Higgs-doublet model (2HDM)~\cite{Gunion:2002zf}. The Higgs sector of DLRSM effectively reduces to a special version of three-Higgs-doublet model at the electroweak scale, leading to this similarity with 2HDM. 

\section{Summary and Outlook}
\label{sec:summary}

We revisit the doublet left-right symmetric model with the scalar sector consisting of a bidoublet and two doublets. The two neutral scalars embedded in the bidoublet and the one from the $SU(2)_L$ doublet can all contribute to EWSB. 
The pattern of the EWSB in this model is parametrized by the two ratios of \textit{vev}s $r = \k_2/\k_1$ and $w = v_L/\k_1$. Most studies of DLRSM take these two ratios to be very small. In this paper, we study how the Higgs data from LHC constrains these two parameters. In particular, we check the consistency of the usual assumption $r, w \ll 1$ with the data.

In our study, we demand that the parameters of our model should be such that
\begin{enumerate}
    \item the Higgs potential is bounded from below,
    \item the quartic couplings of the Higgs potential should satisfy perturbative unitarity,
    \item the Yukawa couplings, particularly of third of generation of fermions, should be perturbative,
    \item the mass-squared values of all CP-even Higgses should be positive.
\end{enumerate}
The third condition in the above list leads to upper limits
$r < 0.8$ and $w < 3.5$, whereas the fourth condition requires the two ratios of quartic couplings, 
\beq
p = \frac{(\a_3-\a_4)}{\a_4} \, \, {\rm and}  \, \,
q = \frac{(\rho_2 - 2 \rho_1)}{2 \rho_1}, \nn
\eeq
to be positive. We vary the parameters of the model, subject to the above constraints and study which values of parameters satisfy the experimental constraints on the lightest CP-even scalar mass $m_h$ and its coupling to gauge bosons $\k_W$ and $\k_Z$, its self-coupling $\k_h$, and its coupling to third generation quarks, $\k_t$ and $\k_b$. We also imposed the additional constraint $m_{H_1} > 15$ TeV.

The model contains ten quartic couplings. In order to get a better understanding of the interplay of different quartic couplings, we defined a simplified set of quartic couplings:
the three quartic couplings $\l_i$, involving only the bidoublet Higgs $\Phi$, are set equal $\l_0$ and three of the quartic couplings $\a_i$, which couple the bidoublet to 
ordinary doublets $\chi_L$ and $\chi_R$, are also set to equal to $\a_0$. A fourth $\l$ coupling is defined in terms of the ratio $x = \l_2/\l_4$, which is constrained to be in the range 
$0.25 \leq x \leq 0.85$ as shown in sec.~\ref{sec:convenient}. This \textit{simple} set contains only six independent parameters, $\{\l_0, \a_0, \rho_1, x, p, q\}$. 

We did our calculations first with the \textit{simple} set of quartic couplings.
We first fixed the value of the quartic coupling ratio $q=1$
and considered four different values for the other quartic coupling ratio $p=0.02, 0.1, 1, 5$. For these values of $p$
and $q$, we picked a random set of values for the other parameters in their allowed ranges. For each random set, we computed $m_h, \k_W, \k_Z, \k_h, \k_t, \k_b$ and $m_{H_1}$ and picked the points which satisfied the constraints on 
all the above quantities. We found that the points which satisfy the $m_h$ constraint, typically satisfied the constraints from $\k_W, \k_Z$, and  $\k_t$ also.
The $\k_h$ constraint rules out a significant number of points allowed by $m_h$ for large values of $w$.
The constraint from $\k_b$ has a non-trivial effect on the allowed points but the $m_{H_1} > 15$ TeV constraint rules out most of the points with low values of $w$. For values of $p \leq 0.1$, there were no allowed points for $w \lesssim 0.6$. 
For  $p>1$, a few random points satisfy all the experimental constraints and have $r< 10^{-2}$ and $w < 10^{-3}$. The number of such points, with very small values of $r$ and $w$, steadily increase with increasing value of $p$ and form $14\%$ of all the allowed random points when $p=5$. We find that such points also require $\rho_1 > 1$. Hence, we can conclude that the usual assumption that $v_L, \k_2 \ll \k_1$ holds true for only a fine-tuned set of parameters, in particular for somewhat large values of the quartic couplings $\rho_1$ and $\rho_2$. This statement is true even when the value of $q$ is moved away from $1$.

We repeated the above calculations with the full, \textit{generic} set of quartic couplings. The low density of allowed points for small values of $r$ and $w$ not only persists in this case but becomes even more pronounced. Hence, our main conclusion
is the following: Most of the studies of the DLRSM assume $\k_1 \approx v$, with other \textit{vev}s being negligibly small. On the contrary, we have pointed out that large values of $r$ and $w$ are allowed by the Higgs data and preferred by the indirect bound on $m_{H1}$. We also note that all the points satisfying $m_{H_1} > 15$ TeV constraint, also satisfy $\k_{t,\, b} \sim 1$ to a very good precision. This confirms that the lightest scalar in DLRSM achieves alignment with the SM-like Higgs through the decoupling of the heavier scalars.

\begin{acknowledgments}
S.K. would like to thank Tuhin S. Roy for valuable discussions. J.M. acknowledges the support of the Department of Science and Technology~(DST) of the Government of India through grant no. SR/WOS-A/PM-6/2019(G). S.K. and S.U.S. thank Ministry of Education, Government of India, for financial support through Institute of Eminence funding.
\end{acknowledgments}

\appendix

\section{Heavy Higgs and gauge boson masses}
\label{appendix:heavyhiggs}

In this section we present the expressions for the masses of the heavy scalars, and the masses and mixing of neutral gauge bosons.  

\noindent $(a)$ Masses of the heavier CP-even neutral Higgses are given by
\bea 
&&m^2_{H_1, H_2} = \frac{1}{4y} (a_1 \mp b_1), \,\,\,\,\,
m^2_{H_3} = 2 \rho_1 v_R^2,  
\eea
Masses of the heavier CP-odd neutral Higgses, $A^0_1$ and $A^0_2$ and the heavier charged scalars are given by
    \begin{equation}
        m^2_{A^0_1, A^0_2} = \frac{1}{4y} (a_2 \mp b_2)\,\, ,  \,\,\,\,
        m^2_{H^{\pm}_1, H^{\pm}_2} = \frac{1}{4y} (a_3 \mp b_3)
    \end{equation}
where 
\bea
        &&x=1+r^2, y=1-r^2,\nn\\
        &&a_1= v_R^2\big(x\alpha_{34}+2\rho_{12}(2w^2+y)\big), \nn\\
        &&  a_2= v_R^2\Big( x\alpha_{34}+2 y\rho_{12}+2 w^2\rho_{12} \Big)+\kappa_1^2\Big(-4xy\lambda_{23}+w^2( x\alpha_{34}+2y\rho_{12})  \Big), \nn\\
        && a_3=v_R^2\Big(x\alpha_{34}+2y\rho_{12}-4w^2\rho_{12}\Big)+\kappa_1^2\Big(2y^2\alpha_{34}+ xw^2(\alpha_{34}-2\rho_{12})\Big), \nn\\
        &&b_1=v_R^2\Bigg[ \Big( x\alpha_{34} +  2\rho_{12} \frac{r^4 + 4 r^2 w^2 -1}{x} \Big)^2 + 16 \rho_{12}^2 r^2 w^2 \frac{(x+w^2)y^2}{x^2} \Bigg]^{1/2}, \nn\\
        &&b_2=\Bigg[v_R^4 \Big( x^2\alpha_{23}^2+4(4r^2w^2-xy)\alpha_{34}\rho_{12}+(xy^2+16r^2w^2(w^2-yr^2))\rho_{12}^2 \Big) \nn\\
         &&+\kappa_1^4\Big(-4xy\lambda_{23}+w^2(x\alpha_{34}-2y\rho_{12})\Big)^2+2\kappa_1^2v_R^2\Big(8r^2w^4\rho_{12}(\alpha_{34}-y\rho_{12})-4xy\lambda_{23}(x\alpha_{34}-2y\rho_{12}) \nn\\
        &&+w^2(x^2\alpha_{34}^2-4y(x\alpha_{34}+8r^2\lambda_{23})\rho_{12}+4y^2\rho_{12}^2)\Big) \Bigg]^{1/2}, \nn\\
      && b_3 = \Bigg[w^4\kappa_1^4(x\alpha_{34}-2y\rho_{12})^2+2v_R^2w^2\kappa_1^2\Big(\alpha_{34}^2(4r^2-y^2)+4(y^2-2r^2y+2r^2w^2)\alpha_{34}\rho_{12} \nn\\ 
      &&  -4y(y^2-r^2y+2r^2w^2)\rho_{12}^2\Big)+v_R^4\Big(x^2\alpha_{34}^2+4(4r^2w^2 - x y)\alpha_{34}\rho_{12}+4(y^2-4r^2y+4r^2w^4)\rho_{12}^2\Big)\Bigg]^{1/2}  \,\,\,,  \nn\\
      \eea 
where $\a_{34}$, $\lambda_{23}$, and $\rho_{12}$ were previously defined. These expressions have been presented in the limit $\m_4 = 0$.

\noindent $(b)$ The massive neutral gauge bosons, $Z^\m_1$ and $Z^\m_2$, and photon $A^\m$ is expressed in terms of the unphysical fields as 
\bea 
A_{\m} &=& \frac{\tilde{g}}{g_L} W^{3}_{L\m} + \frac{\tilde{g}}{g_R} W^{3}_{R\m} + \frac{\tilde{g}}{\gbl} B_{\m},\nn\\
Z_{1\m} &=& \frac{1}{D_1} \Big(\frac{g_L (8 m_{Z_1}^2 - 2 g_R^2 v_R^2)}{(g_R^2 v_R^2 - g_L^2 v_L^2) } W^3_{L\m}+\frac{g_R (8 m_{Z_2}^2 - 2(g_L^2 + g_R^2) \k_+^2 - 2 g_R^2 v_R^2)}{(g_R^2 v_R^2 - g_L^2 v_L^2) } W^3_{R\m} + 2 \gbl B_{\m} \Big),\nn\\
Z_{2\m} &=& \frac{1}{D_2} \Big(\frac{-g_L (8 m_{Z_2}^2 - 2 g_R v_R^2 - 2 \gbl^2 (v_L^2 + v_R^2))}{(g_L^2 v_L^2 - g_R^2 v_R^2)} W^3_{L\mu} \nn\\
&&\,\,\,\,\,\,\,\,\,\,\,\,\,\,\,+ \frac{g_R(8 m_{Z_2}^2 - 2 g_L v_L^2 - 2 \gbl^2 (v_L^2 + v_R^2)}{(g_L^2 v_L^2 - g_R^2 v_R^2)} W^3_{R\m} + 2 \gbl B_{\m} \Big)\,\,\,\,,
\label{eq:zeigenvec}
\eea 
where,
\bea 
\frac{1}{\tilde{g}^2} &=& \frac{1}{g_L^2} + \frac{1}{g_R^2} + \frac{1}{\gbl^2}\,\,\,, \nn\\
D_1 &=& \Big[ 4 \gbl^2 + \frac{1}{(g_L^2 v_L^2 - g_R^2 v_R^2)^2} \Big( 8 m_{Z_2}^2 (g_L^2 + g_R^2) - 4 g_L^2 g_R^2 \k_+^2 - 2 g_L^4 v^2 - 2 g_R^4 V^2  \Big) \Big]^{1/2}\,\,\,, \nn\\
D_2 &=& \Big[ 4 \gbl^2 + \frac{1}{(g_L^2 v_L^2 - g_R^2 v_R^2)^2} \Big((g_L^2 + g_R^2)(8 m_{Z_2}^2 - 2 \gbl^2 (v_L^2 + v_R^2)) - 2 g_L^2 g_R^2 (v_L^2 +v_R^2)  \Big)\Big]^{1/2}\,\,.\nn\\
\eea

\section{Conditions for boundedness from below}
\label{appendix:stability}

In the presence of the most general doublet-bidoublet mixing, here we calculate the conditions for  boundedness from below following the gauge orbit formulation of ref.~\cite{Chauhan:2019fji,Bonilla:2015eha}. The unique form of the quartic term multiplied by $\a_2$ prohibits any simple modification of existing bounds. The quartic potential of our interest is given by eq.~\eqref{eq:scalarpotential}. We choose the following parametrization:
\begin{center}
    $\text{Tr} (\Phi^\dagger \Phi)  + (\chi_L^\dagger \chi_L) + (\chi_R^\dagger \chi_R) \equiv R^2$ \\
    $\text{Tr} (\tilde{\Phi}^\dagger \tilde{\Phi}) = \text{Tr} (\Phi^\dagger \Phi) \equiv R^2 \cos^2 \theta$ \\
    $(\chi_L^\dagger \chi_L) \equiv R^2 \sin^2 \theta \sin^2 \gamma$ \\
    $(\chi_R^\dagger \chi_R) \equiv R^2 \sin^2 \theta \cos^2 \gamma$ \\
    $\text{Tr} (\tilde{\Phi}^\dagger \Phi) / \text{Tr} (\Phi^\dagger \Phi) \equiv \beta e^{-i\omega}$ \\
    $\text{Tr} (\Phi^\dagger \tilde{\Phi}) / \text{Tr} (\Phi^\dagger \Phi) \equiv \beta e^{i\omega}$ \\
    $(\chi_L^\dagger \Phi \Phi^\dagger \chi_L) / \text{Tr} (\Phi^\dagger \Phi) (\chi_L^\dagger \chi_L) \equiv \eta_1$ \\
    $(\chi_R^\dagger \Phi^\dagger \Phi \chi_R)/ \text{Tr} (\Phi^\dagger \Phi) (\chi_R^\dagger \chi_R) \equiv \eta_2$\\
    $(\chi_L^\dagger \tilde{\Phi} \tilde{\Phi}^\dagger \chi_L)/\text{Tr} (\tilde{\Phi}^\dagger \tilde{\Phi}) (\chi_L^\dagger \chi_L) = \eta_3 $\\
    $(\chi_R^\dagger \tilde{\Phi}^\dagger \tilde{\Phi} \chi_R)/ \text{Tr} (\tilde{\Phi}^\dagger \tilde{\Phi}) (\chi_R^\dagger \chi_R) = \eta_4\, ,$
\label{eq:polyparameterisation}
\end{center}
where $R > 0$,  $ \{\theta, \gamma\} \in [0,\frac{\pi}{2}]$, $\beta \in [0,1]$, and $\omega \in [0,2\pi]$. 
The allowed range for these  parameters  is $\frac{1}{2} (1-\sqrt{1-\beta^2}) < \eta_{1,2,3,4} < \frac{1}{2} (1+\sqrt{1-\beta^2})$ along with
\bea
\eta_1 + \eta_3 = 1, \,\,\,\,\, \eta_2 + \eta_4 =1. 
\label{eq:dependency}
\eea

Using the parametric forms defined above we  obtain
\begin{equation}
    V_4 = R^4 \left [\cos^4 \theta \, f_1(\lambda_i, \beta,\omega) + \sin^4 \theta  \, f_2(\rho_i,\gamma) + \sin^2 \theta \cos^2 \theta \, f_3(\alpha_i, \gamma, \eta_i)\right ] \,\,\, ,
\label{eq:BFBV4}
\end{equation}
where 
\begin{eqnarray}
   && f_1(\lambda_i, \beta,\omega)\equiv \frac{1}{4}(\lambda_1 + \lambda_2 \beta^2\cos{2\omega}+\lambda_3 \beta^2+\lambda_4 \beta\cos{\omega}), \nn\\
   && f_2(\rho_i,\gamma) \equiv \rho_1 \sin^4 \gamma + \rho_1 \cos^4 \gamma + \rho_2 \sin^2 \gamma \cos^2 \gamma, \nonumber \\
    &&f_3(\alpha_i, \gamma, \eta_i) \equiv \alpha_1 + 2\alpha_2 \beta\cos{\omega}+\alpha_3 (\eta_1 \sin^2 \gamma + \eta_2 \cos^2 \gamma) + \alpha_4(\eta_3 \sin^2 \gamma + \eta_4 \cos^2 \gamma). \nn
\end{eqnarray}
Upon applying the copositivity criteria, eq.~\eqref{eq:BFBV4} indicates that
\begin{eqnarray}
    &&f_1(\lambda_i, \beta,\omega) > 0, \\
    &&f_2(\rho_i, \gamma) > 0,  \\
    &&f_3(\alpha_i, \gamma, \eta_i)+ 2 \sqrt{f_1(\lambda_i, \beta,\omega)f_2(\rho_i, \gamma)} > 0\,\,\, .
    \label{eq:BFBV4fgh}
\end{eqnarray}
These conditions should hold for all possible values of the parameters $\{\beta,\omega,r,\eta_{1,2,3,4},\gamma\}$.

\noindent $\bullet$ $f_1(\lambda_i, \beta,\omega)$ is minimized at 
$\beta=0$, or $\sin{\omega}=0$, and $\cos \omega =  -\l_4/2\l_2$. Subsequently, certain combinations  of $\l$\,s can be obtained~\cite{Chauhan:2019fji}
\bea
 f_1 > 0:   
 \left\{\begin{array}{lr}
        \lambda_{1}\\
         \left(\lambda_1 -\frac{\lambda_4^2}{2\lambda_2+\lambda_3}\right) \qquad \quad \: \impliedby
 2\lambda_2+\lambda_3>|\lambda_4| \\
        \left(\lambda_{1} +\lambda_{3}
+2( \lambda_{2}
-|\lambda_{4}|)\right) \\
\left(\lambda_{1} +\lambda_{3} -2 \lambda_{2}
-\frac{\lambda_4^2}{4\lambda_2 }\right) \impliedby |4\lambda_2|>|\lambda_4|  \\
        \end{array}\right\}
        \label{eq:bfblambda}
\eea
Here the notation ``p $\impliedby$ q" implies condition p has to checked if and only if condition q is true.

\noindent $\bullet$ $f_2(\rho_i, \gamma)$ is symmetric under the exchange  $\sin \gamma \leftrightarrow \cos \gamma$. Thus $f_2$ becomes minimum  at $\sin \gamma = \cos \gamma =\frac{1}{\sqrt{2}}$ and assumes the value $f_{2,\text{min}} \equiv \frac{2\rho_1 + \rho_2}{4}$. In the limits $\tan \g = 0$, and $\tan \gamma = \infty$, $f_{2,\text{min}} = \rho_1$. This leads to the following boundedness criteria
\bea 
f_2 > 0 : \,\,\,\,\,\, \{ \rho_1 >0, \,\,\,\,\, 2 \rho_1 + \rho_2 > 0 \}\,\,\,\, .
\label{eq:bfbrho}
\eea

\noindent $\bullet$ Eq.~\eqref{eq:BFBV4fgh} is symmetric under the exchange  $\eta_1 \leftrightarrow \eta_2, \eta_3 \leftrightarrow \eta_4$,  along with $\sin \gamma \leftrightarrow \cos \gamma$. 
The LHS of eq.~\eqref{eq:BFBV4fgh} thus takes a minimum value at $\eta_1 = \eta_2$, $\eta_3 = \eta_4$ and $\sin \gamma = \cos \gamma = \frac{1}{\sqrt{2}}$,
\begin{equation}
    \alpha_1 + 2\alpha_2 \beta\cos{\omega}+ \alpha_3 \eta_1 + \alpha_4 \eta_3 + 2\sqrt{f_1(\lambda_i, \beta,\omega) \left(\frac{2\rho_1 + \rho_2}{4}\right)} > 0
\end{equation}
Eq.~\eqref{eq:dependency} implies that $\eta_{1(2)}^{min}=\eta_{3(4)}^{max}$ and vice versa. So third condition becomes
\bea
&&\text{for}\,\,\tan \g = 1 :\nn\\ 
&&   \alpha_1 + 2\alpha_2 \beta\cos{\omega}+ \frac{\alpha_3}{2} (1\pm \sqrt{1-\beta^2})  + \frac{\alpha_4}{2} (1\mp \sqrt{1-\beta^2}) + \sqrt{f_1(\lambda_i, \beta,\omega) \left(\frac{2\rho_1 + \rho_2}{4}\right)} > 0 ,\nn\\
\label{1st}
\eea
\bea
&&\text{for}\,\,\tan \g = 0 \,\,\text{or}\,\, \tan \g =\infty : \nn\\
&&  \alpha_1 + 2\alpha_2 \beta\cos{\omega}+ \frac{\alpha_3}{2} (1\pm \sqrt{1-\beta^2})  + \frac{\alpha_4}{2} (1\mp \sqrt{1-\beta^2}) + \sqrt{f_1(\lambda_i, \beta,\omega)\rho_1} > 0\,\,. \nn\\
\label{2nd}
\eea
The  contribution from $\a_2$ dependent term of $V_4$ in eqs.~\eqref{1st} and \eqref{2nd} is given by $\tilde{f_3}(\alpha_2,\beta,\omega)=2\alpha_2 \beta\cos{\omega}$ which has a minimum value at $\beta=0$ or $\sin{\omega}=0$.

\noindent \textbf{Case 1:} For $\beta=0$,  $\tilde{f_3}(\alpha_2,\beta,\omega)=0$.\\
\textbf{Case 2:} For $ \sin{\omega}=0, \,\, \cos{\omega}=\pm 1$, which implies  $\tilde{f_3}(\alpha_2,\beta,\omega)=-2|\alpha_2|\beta$ \\
\textbf{Case 3:} On the boundary $\beta=1$, $\tilde{f_3}(\alpha_2,\beta,\omega)=-2|\alpha_2|$.

Now we list the boundedness conditions emerging from eqs.~\eqref{1st} and \eqref{2nd} in the aforementioned limits.
\begin{itemize}
    \item For  $\beta=0 $,
    \bea
 \alpha_1 + \alpha_3 +  \sqrt{\lambda_1(2\rho_{1}+\rho_2)}>0 \nn\\
  \alpha_1 + \alpha_4 +  \sqrt{\lambda_1(2\rho_{1}+\rho_2)}>0 \nn\\
 \alpha_1 + \alpha_3 + 2 \sqrt{\lambda_1 \rho_{1}}>0 \nn\\
 \alpha_1 + \alpha_4 +  2\sqrt{\lambda_1 \rho_{1}}>0 \,\,\,.
 \label{eq:bfbalpha1}
 \eea
     \item For  $\sin{\omega=0}$,
\bea    
    &&\alpha_1 -2|\alpha_2|\frac{|\lambda_4|}{2\lambda_2 +\lambda_3}+ \frac{\alpha_3}{2}\left( 1\pm\sqrt{1-\frac{\lambda_4^2}{(2\lambda_2+\lambda_3)^2}}\right) +\frac{\alpha_4}{2}\left( 1\mp \sqrt{1-\frac{\lambda_4^2}{(2\lambda_2+\lambda_3)^2}}\right) \nn\\
     &&+ \sqrt{\left(\lambda_1 -\frac{\lambda_4^2}{2\lambda_2+\lambda_3}\right) (2\rho_{1}+\rho_2)}>0 \,\,\, ,\nn\\
     &&\alpha_1 -2|\alpha_2|\frac{|\lambda_4|}{2\lambda_2 +\lambda_3}+ \frac{\alpha_3}{2}\left( 1\pm\sqrt{1-\frac{\lambda_4^2}{(2\lambda_2+\lambda_3)^2}}\right) +\frac{\alpha_4}{2}\left( 1\mp \sqrt{1-\frac{\lambda_4^2}{(2\lambda_2+\lambda_3)^2}}\right) \nn\\
&&     +\sqrt{\left(\lambda_1 -\frac{\lambda_4^2}{2\lambda_2+\lambda_3}\right)\rho_{1}}>0 \,\,\, .
\label{eq:bfbalpha2}
\eea     
     
      \item For $\beta=1$, $\sin{\omega}=0$,
     \bea     
      &&\alpha_1 -2|\alpha_2|+\frac{\alpha_3}{2}+\frac{\alpha_4}{2} +  \sqrt{\left(\lambda_{1} +\lambda_{3}
  +2( \lambda_{2}
 -|\lambda_{4}| )\right)(2\rho_{1}+\rho_2)}>0\,\, , \nn\\
  && \alpha_1 -2|\alpha_2|+\frac{\alpha_3}{2}+\frac{\alpha_4}{2} +  2\sqrt{\left(\lambda_{1} +\lambda_{3}
  +2( \lambda_{2}
 -|\lambda_{4}| )\right)\rho_{1}}>0\,\,\, . 
 \label{eq:bfbalpha3}
 \eea
 \item For $\beta=1$, $\cos{\omega}=-\lambda_4/2\lambda_2$,
 \bea
 && \alpha_1 -2|\alpha_2|+\frac{\alpha_3}{2}+\frac{\alpha_4}{2} +  \sqrt{\left( \lambda_{1} +\lambda_{3}
  -2 \lambda_{2}
  -\frac{\lambda_{4}^2}{4\lambda_2}\right)(2\rho_{1}+\rho_2)}>0\,\,\, , \nn\\
  && \alpha_1 -2|\alpha_2|+\frac{\alpha_3}{2}+\frac{\alpha_4}{2} +  2\sqrt{\left( \lambda_{1} +\lambda_{3}
  -2 \lambda_{2}
  -\frac{\lambda_{4}^2}{4\lambda_2}\right)\rho_{1}}>0\,\,\, .
  \label{eq:bfbalpha4}
  \eea
\end{itemize}
Eqs.~\eqref{eq:bfblambda}, \eqref{eq:bfbrho}, \eqref{eq:bfbalpha1}, \eqref{eq:bfbalpha2}, \eqref{eq:bfbalpha3}, and \eqref{eq:bfbalpha4} together constitute the full set boundedness from below conditions in DLRSM.

\end{document}